\documentclass[a4paper,fleqn]{cas-sc}
\usepackage[numbers]{natbib}
\usepackage{float}
\usepackage{placeins}
\usepackage{booktabs, threeparttable, array, graphicx}
\usepackage{makecell}
\usepackage{pdfpages}
\usepackage{xcolor}
\usepackage[utf8]{inputenc}
\usepackage{tabularx}   % For adjustable width columns
\usepackage{geometry}
\usepackage[most]{tcolorbox}
\usepackage[utf8]{inputenc}
\newtcolorbox{myframe}[1][]{
  width=0.9\linewidth,
  title={#1},
}
\def\tsc#1{\csdef{#1}{\textsc{\lowercase{#1}}\xspace}}
\tsc{WGM}
\tsc{QE}
\begin{document}
\let\WriteBookmarks\relax
\def\floatpagepagefraction{1}
\def\textpagefraction{.001}

% Short title
\shorttitle{Requirements Perception Gap across Stakeholders of Aged Care Digital Health}    

% Short author
\shortauthors{Xiao et al.}  

% Main title of the paper
\title [mode = title]{Requirements Perception Gap across Stakeholders: A Comparative Survey of Aged Care Digital Health Software}% with Practitioner, Health Providers and Senior End User for Aged-Care Digital Health Apps.}  

% Title footnote mark
% eg: \tnotemark[1]
\tnotemark[1] 

% Title footnote 1.
% eg: \tnotetext[1]{Title footnote text}
\tnotetext[1]{} 

% First author
%
% Options: Use if required
% eg: \author[1,3]{Author Name}[type=editor,
%       style=chinese,
%       auid=000,
%       bioid=1,
%       prefix=Sir,
%       orcid=0000-0000-0000-0000,
%       facebook=<facebook id>,
%       twitter=<twitter id>,
%       linkedin=<linkedin id>,
%       gplus=<gplus id>]

\author[inst1]{Yuqing Xiao}  
% Corresponding author indication
\cormark[1]

% Footnote of the first author
\fnmark[1]
\cortext[1]{Corresponding author}

% Email id of the first author
\ead{}

% URL of the first author
\ead[url]{}

% Credit authorship
% eg: \credit{Conceptualization of this study, Methodology, Software}
\credit{Study design, Ethics, Conceptualisation, Methodology, Data
collection, Data analysis, Original draft preparation, Figures, and
Tables, Revising, Reviewing and Editing}

\author[inst1]{John Grundy}

\affiliation[inst1]{organization={Department of Software Systems and Cybersecurity, Faculty of IT, Monash University},%Department and Organization
            city={Clayton},
            postcode={3800}, 
            state={VIC},
            country={Australia  }}

% Footnote of the second author
\fnmark[2]

% Email id of the second author
\ead{}

% URL of the second author
\ead[url]{}

% Credit authorship
\credit{Study design, Supervision,
Ethics, Reviewing and editing}
\author[inst2]{Anuradha Madugalla}

\affiliation[inst2]{organization={School of IT, Deakin University},%Department and Organization
            city={Melbourne},
            postcode={3125}, 
            state={VIC},
            country={Australia}}
% Footnote of the third author
\fnmark[3]

% Email id of the third author
\ead{}

% URL of the third author
\ead[url]{}
\credit{Supervision, Ethics, Reviewing and Editing}
\author[inst3]{Elizabeth Manias}

\affiliation[inst3]{organization={Department of Medicine, Nursing and Health Sciences, Monash University},%Department and Organization
            city={Clayton},
            postcode={3800}, 
            state={VIC},
            country={Australia  }}
% Footnote of the forth author
\fnmark[4]

% Email id of the forth author
\ead{}

% URL of the forth author
\ead[url]{}
\credit{Study design, Supervision, Ethics, Reviewing and editing}
% Address/affiliation
% Corresponding author text

% Footnote text
\fntext[1]{}

% For a title note without a number/mark
%\nonumnote{}

% Here goes the abstract
\begin{abstract}
 \noindent Context: While the demand for aged-care digital health solutions is rising, the effectiveness of these systems depends on the successful elicitation of requirements from diverse stakeholders. Misalignment between developers’ technical assumptions and the actual requirements of older adults and caregivers—the primary stakeholders—can lead to low software adoption and usability failures.
 
 \noindent Objective: We wanted to explore and compare the perspectives of three key stakeholder groups -- older adults, caregivers (formal health providers and informal caregivers), and Digital Health software developers -- on key functional and non-functional requirements.
 
 \noindent Method: We conducted a survey, designed based on the findings from an existing Systematic Review, to gather and analyse data related to the three stakeholder group's (dis)satisfaction with current aged care digital health software, and their views on key future aged care software requirements. A mixed-methods survey approach was employed, integrating quantitative questionnaire data and qualitative open-ended responses from a total sample of ($N=249$), comprised older adults ($n=103$), formal and informal caregivers ($n=41$), and software developers ($n=105$). Data analysis employed a mixed-methods approach, utilising inferential statistics to compare group satisfaction levels and thematic analysis for qualitative open-ended responses.
 
 \noindent Results: Our analysis reveals a significant ``Requirements Gap." Software developers tended to prioritise advanced features and functional requirements, significantly overestimating user satisfaction with core NFRs such as ease of use and responsiveness. Specifically, while 70\% of developers and 80\% of caregivers expressed satisfaction with current NFRs, only 49\% of older adults agreed. Conversely, developers were more critical of existing functional features (20\% satisfaction) compared to older adults (60\%) and caregivers (65\%), who prioritised simplicity and reliability over feature density.
 
  \noindent Conclusion: We validated the functional and non-functional requirements reported in a prior systematic review and extended them with a large open-ended survey. By combining quantitative and qualitative analysis, we identified where stakeholder priorities align and where they diverge across functional requirements and non-functional requirements in both the current design they used and the future design they desire. Our findings present a stakeholder-gap analysis that can guide future co-design processes with requirements, near-term product decisions, and privacy-by-design recommendations in aged-care digital health.
\end{abstract}

% Use if graphical abstract is present
%\begin{graphicalabstract}
%\includegraphics{}
%\end{graphicalabstract}

% Research highlights
% \begin{highlights}
% \item 
% \item 
% \item 
% \end{highlights}

% Keywords
% Each keyword is seperated by \sep

\maketitle

% Main text
\section{Introduction}
Digital health systems play an
increasingly vital role in supporting both physical and mental
well-being, offering innovative approaches for prevention, monitoring,
and management of health conditions among diverse populations ~\cite{niyomyart_current_2024}. Despite their rapid expansion and demonstrated benefits, effectively identifying and addressing the diverse requirements of all stakeholders remains a persistent challenge ~\cite{ramaswamy_user-centered_2023}. Compared to general digital health solutions, those designed for older adults demand greater attention to stakeholder diversity, as successful aged-care technologies depend not only on the older adults who use them but also on the caregivers who support them and the developers who create them in practice ~\cite{huang_understanding_2025, yang_digital_2024}. Understanding and integrating these perspectives is essential to ensure optimal adoption, user experience(UX), and clinically meaningful outcomes ~\cite{shen_perspectives_2025}. 

In understanding user perspectives, requirements are fundamental to developing any usable and effective technology ~\cite{niyomyart_current_2024}. While much prior research has focused on barriers and challenges---such as usability, accessibility, and trust---these perspectives primarily describe what hinders adoption rather than what enables success ~\cite{giebel_problems_2025,omboni_telehealth_2020, shamsujjoha_developer_2024}. Requirements go beyond identifying obstacles and move into specifying what a system must deliver and how it should perform to meet user expectations and clinical goals ~\cite{gani_understanding_2025, jacob_ai_2025}. Requirements for any system are twofold: functional and non-functional. In relation to aged care digital health applications, ``functional requirements'' define the concrete services the health application must deliver, e.g., login, medication diary, cognitive exercises. ``Non-functional requirements'' refer to the quality attributes that shape how those services are experienced by the user, such as security, user interface, and technical platform ~\cite{trinh_understanding_2023, chen_digital_2023, dong_effectiveness_2023, xiao_requirements_2025}. A comprehensive alignment across both functional and non-functional dimensions ensures that software systems are not only technically robust but also demonstrate high operational fitness and perceived trustworthiness in situated contexts ~\cite{chadwick_engagement_2024}. For instance, a medication reminder’s Functional Requirement (FR)—its scheduling logic—is critical for system reliability, while its Non-Functional Requirement (NFR)—data security and interface intuitiveness—is foundational to user trust.

In the domain of aged-care digital health, the requirements engineering process must account for a complex hierarchy of stakeholders: older adults, caregivers, and software developers ~\cite{bertolazzi_barriers_2024, zainal_exploring_2025}. The relationship between older adults and their caregivers is particularly significant for Requirements Elicitation. Caregivers (both formal and informal) function as Domain Experts and Proxy Users who mediate the interaction between the end-user and the system ~\cite{andrews_older_2019,white_gamification_2022}. Their involvement in Requirements Analysis is essential to uncovering how stakeholder expectations converge or diverge ~\cite{singh_real-world_2024}. While shared priorities often include NFRs like usability and accessibility, caregivers frequently introduce distinct FRs related to monitoring and workflow coordination ~\cite{huang_understanding_2025,bertolazzi_barriers_2024}. Capturing these multi-faceted perspectives leads to more inclusive design and reduces the risk of requirement volatility during the development lifecycle ~\cite{white_gamification_2022,chen_digital_2023}. Software developers represent the third critical stakeholder group. While their technical expertise facilitates the implementation of emerging technologies ~\cite{shamsujjoha_developer_2024}, there is a documented Cognitive Gap between developers and older adult users due to generational and technological experiential differences ~\cite{gani_understanding_2025, jacob_ai_2025}. Practitioners often prioritise Performance Optimisation, Architectural Complexity, and Aesthetic Appeal, whereas end-users and caregivers emphasise NFRs such as Reliability, Cognitive Load Reduction, and Emotional Reassurance ~\cite{niyomyart_current_2024,chen_digital_2023}. Empirical evidence suggests that these divergent mental models often result in systems that satisfy Technical Specifications but fail to meet User-Centred Quality Attributes ~\cite{shamsujjoha_developer_2024,jacob_ai_2025}. This is frequently exacerbated by developers’ limited direct exposure to end-users, leading to an "Ivory Tower" approach to design ~\cite{shamsujjoha_developer_2024, jacob_ai_2025}. By systematically comparing how developers, caregivers, and older adults prioritise FRs and NFRs, this research identifies the Requirements Gap that can be mitigated through Co-design and Participatory Requirements Engineering ~\cite{hepburn_barriers_nodate, ardito_end_2012}.

Despite advancements in human-centric SE, two primary research gaps persist:
\begin{itemize}
    \item  Multi-Stakeholder Convergence: Existing literature lacks empirical studies that compare requirement prioritisation across the triad of older adults, caregivers, and developers ~\cite{shamsujjoha_developer_2024, asl_potential_2023, rathnayake_family_2019}. This limits our understanding of how to synthesise a balanced, representative set of Design Criteria.
    \item Actionable Design Artefacts: Current research focus is heavily skewed toward identifying barriers rather than providing Actionable Guidance for system improvement ~\cite{white_gamification_2022, chan_comparative_2023, hepburn_barriers_nodate}. There is a lack of empirical frameworks that translate stakeholder-defined NFRs into specific Implementation Guidelines to enhance software inclusiveness and trust.

\end{itemize} 

By systematically mapping Current vs. Desired Requirements, this study generates practical insights to inform Requirements Specification, guide developers in Architecture Design, and support sustainable adoption within ageing populations ~\cite{huang_understanding_2025, hepburn_barriers_nodate, abdullah_overview_2021, stamm_exergames_2019}. The interdependent relationship between older adults and their caregivers plays a particularly significant role. Caregivers serve as both indirect users and domain experts who mediate interaction with technology \cite{white_gamification_2022, chen_digital_2023}. Incorporating these complementary perspectives can lead to designs that are more intuitive and responsive to the complex dynamics of aged care---ultimately improving both user engagement and health outcomes \cite{white_gamification_2022, chen_digital_2023}.

\section{Related Work}
\label{sec:related-work}

\subsection{User-centred Requirements Engineering with Multi-stakeholders}
\label{subsec:rw-usercentred-multistakeholder}

Requirements Engineering (RE) is a multi-stakeholder activity because requirements originate from, and must be validated against, the real needs of different stakeholders who use or indirectly use the system. Standards for requirements processes explicitly emphasise the identification of stakeholders and the analysis and documentation of stakeholder requirements as foundational outcomes of RE \cite{xiao_requirements_2025, pacheco2012systematic,ramaswamy_user-centered_2023}. This multi-stakeholder nature is critical for empirical SE or RE research, and existing studies have shown that incorrect or incomplete stakeholder identification is associated with inadequate requirements engineering and downstream project risks, motivating systematic approaches to stakeholder identification and involvement \cite{muller_so_2022,noauthor_stakeholders_nodate,elkady_prioritizing_2024}. Furthermore, in recent development contexts (e.g., agile and large-scale settings), stakeholders' involvement is further framed as essential for maintaining feedback loops that keep requirements aligned with evolving needs in each delivery~\cite{schon2017agile}.

For digital health (DH) software, RE requires extending ``user"-centred approaches to include multiple stakeholder groups with different roles. For example, in aged care digital health software, developers, older adults, and caregivers (both informal and formal) often evaluate system value differently, and literature has identified gaps between developer assumptions and the lived needs of ageing users and care stakeholders~\cite{xiao_requirements_2025}. Hence, multi-stakeholder engagement is frequently advocated as a way to reduce requirement misunderstandings and to improve fit with real-world practices and constraints.

Engaging multiple stakeholders can provide several RE benefits. First, it can help capture requirements from different perspectives rather than being limited to a single group (e.g., in DH software, health providers can reveal the clinical requirements of older adults, while seniors can report their own accessibility requirements, and developers might focus on the technical requirements), thereby improving requirements completeness and relevance~\cite{xiao_requirements_2025, shamsujjoha_developer_2024,ardito_end_2012,an_older_2024}. Second, it supports early surfacing of conflicts in requirements (e.g., deaf users' requirements and their caregivers' requirements in a home alarm system), enabling negotiation and prioritisation before implementation \cite{xiao_requirements_2025, kuusik_home_2012,jussli_senior_2021}. Third, it strengthens and validates RE by verifying whether the requirements statements and proposed designs actually satisfy stakeholder requirements, by broadening the set of perspectives, consistent with the intent of stakeholder-needs-driven requirements processes~\cite{xiao_requirements_2025, scandurra_user_2008, tran-nguyen_mobile_2022, shamsujjoha_developer_2024}. Prior user-centred work has operationalised multi-stakeholder engagement through user studies, where older adults and other stakeholders contribute throughout design activities rather than only at evaluation time. For example, surveys, interviews, and focus groups for ageing in place synthesise evidence on co-design approaches, barriers, and facilitators \cite{ardito_end_2012,le_design_2014,nielsen_user-innovated_2018,ariaeinejad_user-centered_2016}. Such work highlights that engagement outcomes are not limited to usability improvements: studies report benefits such as better contextual fit, improved acceptability, and clearer understanding of stakeholder priorities and constraints, while also documenting practical challenges (e.g., recruitment, power imbalances, and sustaining participation) that directly affect the quality of elicited requirements \cite{cleland_contextualizing_2015,wu_acceptability_2024,lopes_redefining_2010, madampe2024struggle}. Together, these findings motivate RE studies that compare perspectives across stakeholder groups to identify alignment, divergence, and actionable requirement implications.

\subsection{Functional and Non-functional Requirements for digital health software}

Functional requirements define system capabilities that stakeholders expect. In aged-care digital health, FRs frequently include monitoring, safety support, medication management, rehabilitation support, communication, and coordination. Developers may judge satisfaction through feasibility and stability, while older adults may judge satisfaction through whether a capability reduces effort, anxiety, or uncertainty in daily life. Caregivers may prioritise functions that reduce burden or improve coordination, even if older adults perceive them differently. A comparative satisfaction--limitation--future framing therefore supports an RE-style contribution by producing evidence that can inform prioritisation and distinguish missing capability areas from underperforming ones. Several studies have focused exclusively on older adults’ needs or preferences, without incorporating the perspectives of other stakeholders. For example, Chan et al. identified that older adults expressed high behavioural intention toward virtual reality (VR) in telerehabilitation, shaped by effort expectancy and trust, but the study excluded caregiver or developer perspectives ~\cite{chan_comparative_2023}. Similarly, Mao et al. conducted a mixed-methods needs assessment in older adult communities to identify barriers to telemedicine, such as lack of technical confidence, difficulty hearing, and language limitations~\cite{mao_barriers_2022}. These works offer important insights into user-reported barriers but do not consider how these perspectives compare with those of developers or other stakeholders such as caregivers.

Non-functional requirements capture quality attributes such as usability, accessibility, reliability, responsiveness, safety, and fit with workflows. In digital health, adoption failures are often explained less by missing features than by quality breakdowns: systems may exist but be difficult to learn, slow, fragmented, or demanding in real contexts of use. This aligns with established quality models and usability standards \cite{esaki2001system,iso2010ergonomics}. Nevertheless, NFRs in health technology are often framed vaguely at a high level (e.g., what are the barriers that stop users using ...'') rather than defined in ways (e.g., as specific NFRs'') that support engineering action (i.e., which NFRs are met, which are failing, and which should be prioritised next)~\cite{mao_barriers_2022}. Stakeholders also operationalise NFR satisfaction differently. For example, research on tele-rehabilitation, telemedicine, and activity monitoring shows that developers may consider an NFR satisfied when it meets standard engineering expectations, while older adults may evaluate it relative to health conditions (vision, hearing, mobility, cognition) and reliance on support networks~\cite{omboni_telehealth_2020,cleland_contextualizing_2015,kuusik_home_2012}. Meanwhile, Mahmoudi Asl et al. investigated stakeholder attitudes (including managers and carers) toward implementing a social robot in dementia care centres and found that caregivers may emphasise robustness, clarity, and routine compatibility because they manage breakdowns and exceptions in everyday care~\cite{asl_potential_2023}. Hence, separating satisfied NFRs, limitations, and future desired NFRs provides a structured basis for stakeholder-aware acceptance criteria and prioritisation.

NFRs for data governance are also important, as they contribute to the privacy and security of the system and ensure trust. Digital health systems are particularly sensitive to data governance, where privacy, security, consent, access control, and compliance requirements influence not only system design but also perceived legitimacy. Many studies measure privacy concerns as attitudes, but fewer separate: (i) methods adopted for privacy-related requirements (controls and practices), (ii) stakeholder opinions about adequacy and trade-offs, and (iii) future desires for privacy and security requirements~\cite{voria2025catalog,gentili2023characterizing, alhammad2024patients}. This separation matters because developers may treat implemented mechanisms as evidence of satisfaction, whereas older adults and caregivers may evaluate privacy/security through trust cues, transparency, and perceived control. Recent mHealth work has proposed structured criteria and tools for assessing privacy and security in apps \cite{rezaee2023critical}, and systematic reviews have synthesised how users perceive confidentiality and security risks \cite{alhammad2024patients}. These studies support treating privacy/security as both NFRs and governance constraints that shape adoption and verification burden.

This leaves a significant research gap in understanding: (1) which functional and non-functional requirements are perceived as most important, (2) whether these requirements are currently supported, and (3) how future visions align or diverge across the three main stakeholder groups.

\subsection{Stakeholder comparisons as an RE contribution: from opinions to prioritisation and trade-offs}
A key contribution of three-group comparison is treating stakeholder perspectives as inputs to RE decisions: what to prioritise, refine, and govern. Comparative designs can surface misalignment (developers perceive a requirement as satisfied while older adults or caregivers disagree), role-specific priorities, and trade-off preferences (e.g., stricter privacy versus easier sharing with caregivers; richer functionality versus simplicity; automation versus transparency). Such structuring is consistent with empirical SE/RE research that uses practitioner-oriented studies to connect perceptions to quality issues and actionable implications~\cite{shamsujjoha_developer_2024, gentili2023characterizing}. For instance, perception-focused studies on trust, transparency, and adoption have demonstrated how stakeholder discourse reveals adoption barriers and risk narratives relevant to SE decision-making \cite{basha2025trust}. Similarly, Balasubramaniam et al. have also emphasised the importance of defining and operationalising transparency and explainability requirements in practice \cite{balasubramaniam2024candidate}.

% % Numbered list
% % Use the style of numbering in square brackets.
% % If nothing is used, default style will be taken.
% %\begin{enumerate}[a)]
% %\item 
% %\item 
% %\item 
% %\end{enumerate}  

% % Unnumbered list
% %\begin{itemize}
% %\item 
% %\item 
% %\item 
% %\end{itemize}  

% % Description list
% %\begin{description}
% %\item[]
% %\item[] 
% %\item[] 
% %\end{description}  

% % %\clearpage %%Remove this from your manuscript

% % Figure
% % \begin{figure}%[]
% %   \centering
% % %    \includegraphics[width=0.9\textwidth]{}
% %     \caption{}\label{fig1}
% % \end{figure}

% % \begin{table}%[]
% % \caption{}\label{tbl1}
% % \begin{tabular*}{\tblwidth}{@{}LL@{}}
% % \toprule
% %   &  \\ % Table header row
% % \midrule
% %  & \\
% %  & \\
% %  & \\
% %  & \\
% % \bottomrule
% % \end{tabular*}
% % \end{table}

\section{Method}
\label{sec:method}

\subsection{Research Questions}
To examine the current and future requirements for aged care digital health apps from multiple stakeholder perspectives, we wanted to answer the following overarching research question:

\begin{quote}
\textit{What are the current and future requirements of aged-care Digital Health Apps from multiple stakeholder perspectives?}
\end{quote}

To address this, the following sub-questions are investigated:

\begin{itemize}
    \item[\textbf{RQ1.}] What requirements are currently perceived as satisfactorily fulfilled?
    \begin{itemize}
        \item[\textbf{RQ1.1.}] What functional requirements are currently perceived as satisfactorily fulfilled?
        \item[\textbf{RQ1.2.}] What non-functional requirements are currently perceived as satisfactorily fulfilled?
    \end{itemize}
    \item[\textbf{RQ2.}] What requirements are currently perceived as unsatisfactorily fulfilled?
    \begin{itemize}
        \item[\textbf{RQ2.1.}] What functional requirements are currently perceived as unsatisfactorily fulfilled?
        \item[\textbf{RQ2.2.}] What non-functional requirements are currently perceived as unsatisfactorily fulfilled?
    \end{itemize}
    \item[\textbf{RQ3.}] What future requirements are desired to support innovation?
    \begin{itemize}
        \item[\textbf{RQ3.1.}] What future functional requirements are desired to support innovation?
        \item[\textbf{RQ3.2.}] What future non-functional requirements are desired to support innovation?
    \end{itemize}
    \item[\textbf{RQ4.}] What perception gaps exist among stakeholder groups regarding current and future requirements?
    \begin{itemize}
        \item[\textbf{RQ4.1.}] What perception gaps exist regarding satisfactorily fulfilled requirements?
        \item[\textbf{RQ4.2.}] What perception gaps exist regarding unsatisfactorily fulfilled requirements?
        \item[\textbf{RQ4.3.}] What perception gaps exist regarding future requirements for innovation?
    \end{itemize}
\end{itemize}

\subsection{Study Design}

To answer our research questions above we conducted a cross-sectional, online, three-group comparative survey ($N=249$) to investigate how older adults ($n=103$), caregivers ($n=41$), and software developers ($n=105$) perceive (i) functional requirements, (ii) general non-functional requirements, and (iii) data governance (privacy/security) requirements for aged-care digital health systems. The survey combined closed-ended Likert-scale items with open-ended questions to capture both structured comparisons and explanatory insights. After data collection and quality checking, we used quantitative methods to analyse the Likert-scale response data and qualitative methods to analyse the open-ended responses. This mixed-methods design enabled us to compare stakeholder perceptions systematically while also capturing the reasoning underlying those perceptions.

 %Functional requirements specify what a system should do, while Non-functional requirements describe how it should perform. 
We used the functional and non-functional requirements identified in our previous systematic literature review (SLR) as the basis for the requirements in our survey\cite{xiao_requirements_2025}. Using these SLR findings, we designed the survey for our target three stakeholder groups. By using identical five-point Likert scales each requirement in all surveys, the study facilitates ease of understanding for our target stakeholders. An a priori power analysis was conducted in ``GPower" to estimate the minimum sample size required for adequate statistical power; larger samples also help improve the practical robustness of parametric analyses to moderate departures from normality~\cite{faul2007g, norman2010likert}.

After completing the internal review of the survey among all authors, a pilot survey was conducted with 8 participants (3 older adults, 3 developers, and 2 caregivers—one formal and one informal). Feedback from this pilot informed revisions to improve clarity, readability, and contextual appropriateness, particularly for older adult respondents, while ensuring that opinions were collected on the same requirements for comparison. Based on the pilot feedback, the wording of technical items was simplified for audiences, redundant or low-relevance questions were deleted, examples were added to illustrate specific features (e.g., medication diary, alert notifications), and the overall layout was reorganised for smoother navigation. Font sizes were enlarged, and formatting was optimised for accessibility across devices, including mobile phones, iPads, and laptops. Pilot responses were retained in the final dataset after anonymisation and removal of items judged irrelevant. All data were then reordered after assigning anonymised unique identifiers. The pilot served as a form of content validation, ensuring that all items were meaningful, comprehensible, and representative of the requirements identified in the SLR. The pilot and final versions of surveys are provided in the Online Supplementary Material. Prior to full deployment, the survey underwent iterative refinement to improve clarity, reduce ambiguity, and ensure the items were interpretable across the three participant groups. Refinement focused on (i) ensuring consistent meaning of requirement terms across stakeholders, (ii) reducing jargon for older adults and caregivers, and (iii) improving the flow and completion time. (If applicable, specify: number of pilot participants, recruitment, and what changes were made.)

\subsection{Survey Questionnaire Design}
In the survey, each version with customised language was developed for each question to suit older adults, caregivers, and developers, thereby capturing the unique perspectives of each group while maintaining methodological comparability across the cohorts. All surveys included questions to collect demographics, older adults' human and health aspects, stakeholders' feedback on current satisfaction or issues with digital health applications, as well as future desirability of features. Open-ended questions were supplemented in each survey to elicit qualitative insights for current acceptance and future expectations. The study received ethics approval prior to recruitment. Participants were provided with study information and consent materials at the start of the survey and could exit at any time.
The survey instrument was organised around three requirement constructs aligned with requirements engineering categories:
\begin{itemize}
  \item \textbf{Functional requirements (FR)}: Core features that define specific behaviour, functions, and user tasks. FRs specify actions, features, and behaviours.
  \item \textbf{Non-functional requirements (NFR)}: Requirements that define how the system performs. NFRs are quality attributes such as security, speed, and usability.  
  % In addition to standard NFRs, we examined a specialised subset --- data governance requirements --- covering privacy and security protections, and related governance expectations.
\end{itemize}
For each construct, we collected participants' opinions on three questions: \emph{currently satisfied}, \emph{current limitations}, and \emph{future desired requirements}.

We used a mixed-methods survey to include both quantitative responses and qualitative open-ended responses. Most closed-ended items used 5-point Likert scales. For example, satisfaction items measured perceived helpfulness on a $5$-point scale (e.g., $1=$ not helpful to $5=$ very helpful), and future-desire items measured desirability on a $5$-point scale (e.g., $1=$ not desired to $5=$ very much desired). We used Likert-scale response formats because they are commonly adopted in SE/RE perception studies to quantify attitudes and facilitate group comparisons. Open-ended responses complemented the Likert items by allowing participants to explain ratings, report concrete issues, and suggest additional requirements not covered by the predefined list. In total, for informal and formal caregivers, there are 24 questions: 16 multiple-choice and 8 open-ended. For older adults, there are 25 questions: 19 multiple-choice and 8 open-ended. Their questionnaire are similar to those of caregivers, but have one more question to answer: how comfortable they are to share data and with whom. For developers, there are 24 questions: 18 multiple-choice and 6 open-ended. Their questionnaire are similar to those of caregivers. The only difference is that they do not answer from a clinical point of view but a technical one. The full version of the survey questions is provided in the Online Supplementary Material.

\subsection{Recruitment and Data Quality Checks}
\begin{figure}
    \centering
    \includegraphics[width=0.6\linewidth]{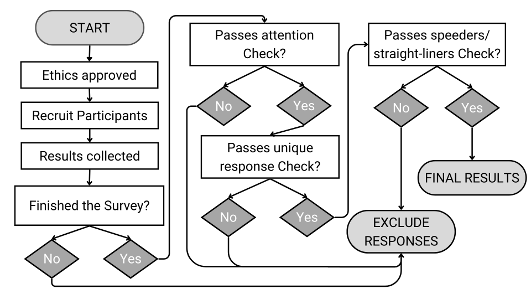}
    \caption{The workflow of recruit, include, and exclude participants for surveys}
    \label{fig:recruitment_workflow}
\end{figure}

We targeted three stakeholder groups. Participants were assigned to a group based on self-reported role:
\begin{itemize}
  \item \textbf{Older adults}: Individuals are older than 65, or self-identified as older adults (The policy can differ in various regions, e.g., in Thailand, 60+; in China, 60+ female), and could complete the survey independently.
  \item \textbf{Caregivers}: individuals providing informal or formal care support to older adults (eg. Formal caregivers include clinicians, nurses, full-time home carers; informal caregivers include family, friends, etc.)
  \item \textbf{Developers}: individuals with software development experience relevant to aged-care digital health technology, or digital health software systems in general (eg. wearable vital signs monitoring software, home care robotics, and in-hospital monitoring software)
\end{itemize}

A large-scale online survey was then deployed using Qualtrics to reach a diverse and representative sample. The survey was self-administered and distributed through social media (e.g., Facebook, LinkedIn), professional networks (e.g., academic mailing lists, digital health company forums), and community digital channels (e.g., seniors online groups) after receiving ethical approval. As figure~\ref{fig:recruitment_workflow} summarizes, after ethics approval, participants were recruited through social media and the professional connections of the authors. 

Data quality of the survey was evaluated based on the following criteria:
\begin{enumerate}
  \item \textbf{Completion check}: We excluded responses that did not reach the end of the survey (i.e., unfinished submissions).
  \item \textbf{Attention check / full-response review}: We included attention-check items and reviewed responses for indications of inattentive answering.
  \item \textbf{Unique response check}: We examined IP address information (where available), demographic patterns, and open-ended responses to identify potential duplicates.
  \item \textbf{Speeding / straight-lining check}: We screened for implausibly short completion times and response patterns indicative of straight-lining.
\end{enumerate}

As figure~\ref{fig:recruitment_workflow} shows, we used this criteria to exclude in four steps: (1) a completion check to ensure surveys were finished; (2) an attention check, which excluded 14 participants; (3) a unique replier check using IP addresses, demographics, and qualitative responses, which found no duplicates; and (4) a quality check to exclude speeder and straight-liner response speed/straight-lining check, which excluded two participants. After the data quality checks, our final analytic sample comprised 249 responses (developers = 105, older adults = 103, caregivers = 41). This staged approach aimed to increase internal validity by minimising low-effort responses and improving the reliability of group comparisons.

\subsection{Data Analysis Method}
\subsubsection{Quantitative analysis}
Quantitative questions used a five-point Likert scale. For descriptive reporting, mean values and 95\% confidence intervals (CI) were calculated to measure overall satisfaction and expectations toward aged-care software features. We report results overall and stratified by stakeholder group (older adults, caregivers, developers). The anonymised raw data and categorised calculations are provided in the Online Supplementary Material. To support three-group comparisons, we compared distributions across groups for each requirement item/category. Because Likert-type data are ordinal, we conducted group comparisons using one-way ANOVA. The combination of descriptive (eg. mean, standard deviation) and inferential statistics allows us to show both the magnitude and precision of group ratings and to test whether the overall differences among groups are greater than would be expected from within-group variability alone~\cite{norman2010likert,sullivan2013analyzing,gardner1986confidence,yaddanapudi2016american,cohen2013statistical}. Although we do not assume all the data fully satisfy the normality assumption, we additionally assessed the magnitude of the observed group differences using effect size estimates to test against non-normal distribution and entered the corresponding values into ``GPower" for effect-size-based interpretation as a complementary analysis~\cite{faul2007g}. Prior methodological work supports the practical use of means and parametric comparisons for such data in many applied research settings~\cite{ho1965robustness,havlicek1974robustness}.  

\subsubsection{Qualitative analysis}
Qualitative data from open-ended survey responses were analysed using Thematic Synthesis, following the guidelines for qualitative research~\cite{thomas_methods_2008, cruzes_recommended_2011}. We employed an iterative three-stage process: (1) initial inductive coding of responses; (2) clustering and categorisation of codes into descriptive sub-themes based on conceptual similarity; and (3) iterative merging and refinement until 'theoretical saturation' was achieved. This ensured that final themes were representative and appropriately granular. The first author conducted all coding, developing an initial codebook through inductive coding of all responses. A subset was independently reviewed by co-authors, with disagreements resolved through consensus discussion until agreement was reached. Thematic saturation was assessed by tracking new codes per tranche of 20 responses; no new codes emerged after approximately the 120th response. To protect participant privacy, qualitative results are reported using anonymised summaries. The coding method and outputs are provided in Appendix~A and the Online Supplementary Material.

\subsubsection{Criteria for identify ``satisfaction", ``limitation", and``desired", with Quantitative and Qualitative results}
Stakeholder satisfaction levels for aged-care software features, along with future desired functional and non-functional requirements, were then identified. We identified key requirements in three outcome categories by triangulating quantitative and qualitative evidence. The interpretation process followed three criteria in each type(current satified, current limited, and future desired): 

\begin{enumerate}
  \item \textbf{Current satisfaction}: a requirement was labelled as currently satisfied if (i) If more than half of participants rated a requirement as ``somewhat helpful” or ``very helpful,” the maximum mean value among groups exceeded 4.5, and qualitative comments were positive.(ii) the maximum mean among the three stakeholder groups exceeded 4.5, and (iii) qualitative responses mentioned it with predominantly positive rather than negative tone.

  \item \textbf{Current limitation}: a requirement was labelled as a  current limitation if (i) the mean Likert value for current satisfaction was low ($\leq 3$) \emph{or} there was a large gap between stakeholder groups, and (ii) qualitative responses reported issues or predominantly negative tone.
  
\item \textbf{Future desired innovation}: a requirement was labelled as future desired if (i) more than half of participants rated it as ``somewhat desired'' or ``very much desired''; (ii) the maximum mean among stakeholder groups exceeded 4.5; and (iii) qualitative responses mentioned it in desired requirements.
\end{enumerate}

This rule-based synthesis was designed to produce transparent, reproducible classifications that connect stakeholder ratings to interpretable requirement outcomes.

\section{Results}\label{results}

We initially received 265 responses: older adults (n = 111), caregivers(n = 41), and developers (n = 113). After checking attention questions and response completeness, the final analytic sample comprised 249 participants (\emph{older adults = 103, caregivers = 41, developers = 105}). Participants were anonymised and assigned coded identifiers according to their group: OA for older adults, FC for formal caregivers, IC for informal caregivers, and Dev for developers (e.g., \textit{OA01},\textit{FC10}, \textit{IC01}, \textit{Dev99}). As the survey was distributed through social media and professional networks rather than a closed mailing list with known membership, it was not possible to determine an overall response rate. Across the open-ended questions, the average response rate was 59\%: older adults (44.2\%), caregivers (61\%), and developers (73.9\%). 

Overall, as summarised in Table~\ref{tab:func_nonfunc_list}, we identified the key functional requirements such as vitals and health monitoring, fall and mobility support, and medication safety, and key non-functional requirements such as simplicity of use, support for cognitive and socio-economic challenges, etc., and key data governance-related issues like data collection and data protection.

\begin{table}[!ht]
\caption{Identified functional and non-functional requirements from the SLR, with additional requirements emerging from survey open-ended responses underlined.}\label{tbl1}
\begin{tabular*}{\tblwidth}{@{}LL@{}}
\toprule
\textbf{Functional Requirements} & \textbf{Non-Functional Requirements} \\ 
\midrule
1. Vitals \& Health Monitoring        & 1. Simple to Use \& Reduce Confusion \\
2. Fall \& Mobility Support           & 2. Health Information \& Tutorials \\
3. Medication \& Safety               & 3. Entertainment \& Social Media \\
4. Brain Training                     & 4. Life Quality Support \\
5. Rehabilitation Support             & 5. Personalisation \\
6. Life Assistance                    & 6. Quick Response \\
7. Chronic Disease Management         & 7. Seamless Integrated \\
8. Fitness Support                    & 8. \underline{Enhance Motivation} \\
9. \underline{Nutrition \& Dietary Management}    & 9. \underline{Support for Cognitive Challenges} \\
10. \underline{Pain Management}       & 10. \underline{Socio-economics Support} \\
                                      & 11. \underline{Attention Support} \\
                                      & 12. \underline{Support from Caregivers} \\
                                      & 13. \underline{Speech \& Language Support} \\
                                      & 14. \underline{Visual accessibility} \\
                                      & 15. \underline{Hearing Support} \\
                                      & 16. \underline{Emotional Support} \\
                                      & \textbf{Data Governance:} \\
                                      & 17. Data Security \\
                                      & 18. Data Privacy \\
                                      & 19. \underline{Data Compliance \& Regulatory} \\
\bottomrule
\end{tabular*}
\label{tab:func_nonfunc_list}
\footnotesize{$^a$ Underlined requirements were newly identified from open-ended survey responses and were not present in the original systematic literature review.}
\end{table}

As summarised in Table~\ref{tab:demographics}, we collected the demographics of older adults' demographics (as themselves, their care recipients, or end users) of age and platform used. The table shows that most older adult participants were concentrated in the younger-seniors groups, especially 60–69 years, while caregivers and developers reported caring for or designing for a somewhat broader age distribution, including more users aged 70 and above. Across all three groups, smartphones were the most commonly used platform, followed by tablets and PCs, whereas smartwatches and physical button phones were reported less frequently, though caregivers noted comparatively higher use of button phones among senior users.
\begin{table*}[!ht]
\caption{The demographics information of Participants.}\label{tab:demographics}
\begin{tabular}{lccc}
\hline
\textbf{Characteristic} & \textbf{Older Adults (n=103)} & \textbf{Caregivers (n=41)} & \textbf{Developers (n=105)} \\
\hline
\multicolumn{4}{l}{\textbf{Age groups of senior users (years), n (\%)}} \\
60--64 & 34 (33.0) & 19 (55.9) & 8 (7.6) \\
65--69 & 32 (31.1) & 20 (58.8) & 34 (32.4) \\
70--74 & 20 (19.5) & 27 (79.4) & 22 (21.0) \\
75--79 & 9 (8.7) & 22 (64.7) & 22 (21.0) \\
80--84 & 4 (3.9) & 20 (58.8) & 15 (14.3) \\
85--89 & 4 (3.9) & 16 (47.1) & 6 (5.7) \\
90--94 & 0 (0.0) & 15 (44.1) & 1 (1.0) \\
$\geq$90 & 0 (0.0) & 10 (29.4) & 0 (0.0) \\
\hline
\multicolumn{4}{l}{\textbf{Platform used by senior users, n (\%)}} \\
Smartphones & 68 (66.0) & 25 (60.9) & 87 (83.7) \\
Physical button phones & 20 (19.0) & 16 (39.0) & 32 (30.8) \\
Smart watch & 17 (16.5) & 5 (21.2) & 56 (53.8) \\
Tablet & 30 (29.1) & 21 (51.2) & 64 (61.5) \\
PC & 47 (45.6) & 10 (24.4) & 50 (48.1) \\
\hline
\end{tabular}
\begin{minipage}{\tblwidth}
\smallskip
$^a$The age group reported by caregivers and developers are the age group of their care recipients or end users; the results they report can include multiple age groups if their work scope involves multiple age groups of older adults.
\end{minipage}
\end{table*}

\subsection{RQ1. What are the currently satisfied requirements}

\begin{figure}
    \centering
    \includegraphics[width=0.875\textwidth]{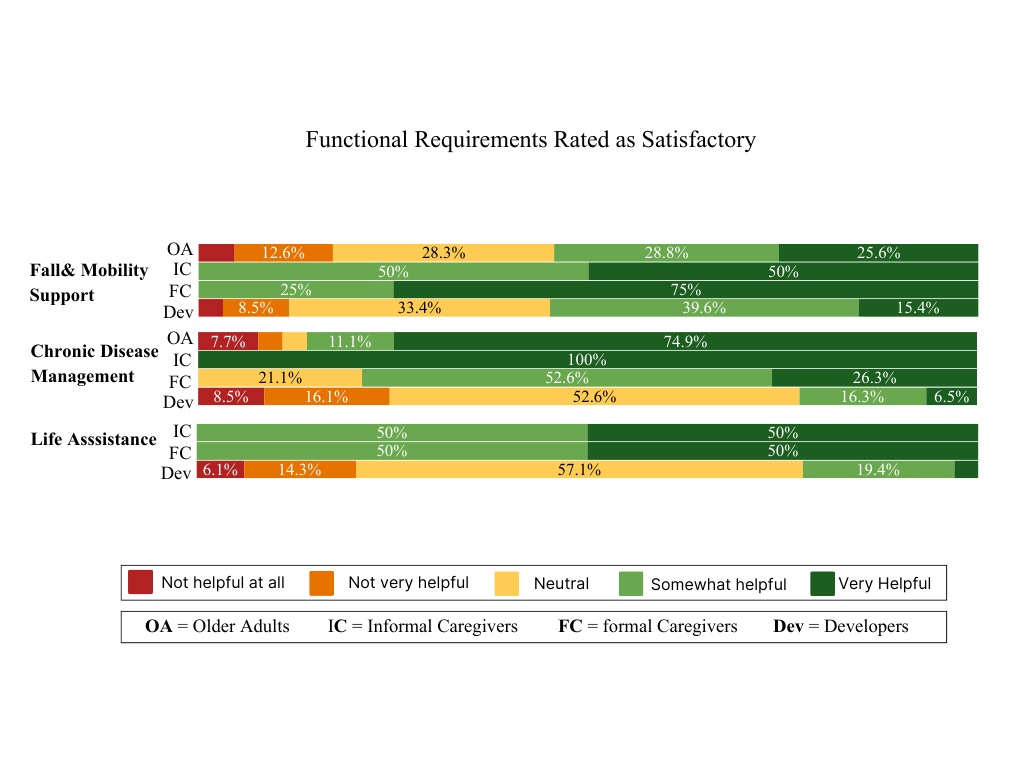}
    \caption{Functional requirements rated as satisfactory: Current satisfaction levels across stakeholder groups, showing proportion distributions.}
    \label{fig:cur_func_satis}
\end{figure}

\begin{figure}
    \centering
    \includegraphics[width=0.875\textwidth]{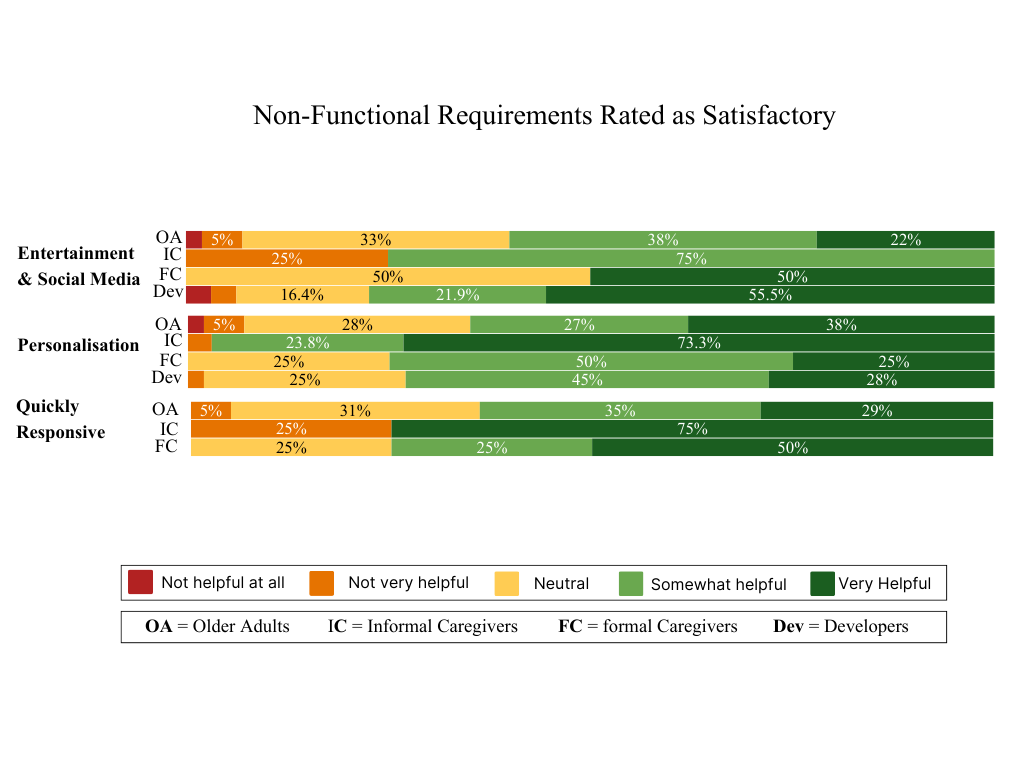}
    \caption{Non-Functional requirements rated as satisfactory: Current satisfaction levels across stakeholder groups, showing proportion distributions.}
    \label{fig:cur_nonfunc_satis}
\end{figure}

Figures~\ref{fig:cur_func_satis} and~\ref{fig:cur_nonfunc_satis} illustrate the distribution of Likert-scale responses reflecting stakeholders’ current satisfaction with functional and non-functional requirements. Each bar represents a stakeholder group—older adult end users, caregivers (formal and informal), and developers—and shows the percentage of responses across the five satisfaction categories, from not helpful at all'' to very helpful''. 

%\clearpage
\subsubsection{Satisfied Functional Requirements}
As shown in Figure~\ref{fig:cur_func_satis}, the most highly rated functional requirements were typically AI-based or conventional digital health features that are familiar to experienced developers, although not necessarily to healthcare providers or older adult users. These functions included vital signs and health monitoring, fall and mobility support, and life assistance for chronic disease management. Stakeholder perceptions followed a broadly similar pattern, indicating overall appreciation but uneven levels of satisfaction across groups. Caregivers, both formal and informal, rated these features most positively, reflecting their perceived usefulness in everyday care. Older adults also viewed them favourably, although with greater variability: over 65\% reported being satisfied, while fewer than 20\% considered them ``not helpful at all.'' By contrast, developers reported the lowest level of satisfaction, with more than half indicating dissatisfaction.

To illustrate these perspectives, participants highlighted the importance of timely monitoring and meaningful data integration. For example, FC1 noted, ``… should observe abnormal changes in time and communicate these vital signs to their family members … It would be best if the family could chat with the elderly …” Similarly, Dev4 emphasized, ``I want to design a reliable monitoring system to track health status,” while OA18 shared, ``I’d like an easy check on my blood oxygen … I can notice changes early … give me peace of mind to go to sleep.” These reflections illustrate how vital sign monitoring serves multiple purposes—enabling early detection of health changes, providing reassurance for both older adults and families, and informing technical integration in system design. Several older adults expressed similar needs, such as one who stated, ``I want something that connects my daily measurements directly to my doctor … otherwise it feels like I’m just writing numbers down with no use.”
\subsubsection{Satisfied Non-Functional Requirements}
Among the non-functional requirements, ``Entertainment and Social Media,” ``Personalisation,” and ``Quick Responsiveness” stood out as the most consistently valued features across stakeholder groups (Figure~\ref{fig:cur_nonfunc_satis}). They exhibited similar patterns on the satisfaction scale, with both older adults and caregivers generally rating them as satisfactory. However, developers expressed lower satisfaction with ``Entertainment and Social Media,” believing that digital health applications should incorporate more engaging and enjoyable elements to enhance user experience. In contrast, older adults and caregivers showed limited interest in entertainment features within digital health software, suggesting a mismatch between developers’ design focus and end-user priorities. Additionally, older adults reported lower satisfaction with ``Personalisation" compared with other participant groups. This indicates that although personalisation was rated positively overall, future designs should further enhance adaptability and individual relevance to help older adults feel that the system truly reflects their personal needs and preferences.
Caregivers emphasised the ``Quickly Responsive” importance for maintaining user trust, whereas developers and older adults identified performance delays as a source of frustration. Older adults reported that slow or inconsistent system feedback discouraged continued use. OA22 explained, ``When I press a button, it should react right away — if it’s slow, I stop trusting it.” FC01 confirmed this sentiment, ``…delayed responses often led older users to believe they had made mistakes.” These experiences indicate that real-time responsiveness is not merely a technical attribute but a core determinant of reliability and confidence in aged-care applications. 

\begin{center}
\begin{myframe}[\centering\textbf{RQ1 Answer Summary}]
\footnotesize
Collectively, these findings suggest that while current digital health systems deliver useful AI-based functionalities with a good amount of user interaction, the current software generally satisfies long-existing, classic digital health functional requirements and well-explored, easy-to-appropriate non-functional requirements best. The lesson we learned here is that \textbf{\textit{practice did make perfect}}. The good amount of existing attempts in digital health design and the established user interaction and engagement requirements (even for other types of apps) heuristically support good products that align with the real-world requirements of older adults and caregivers.
\end{myframe}
\end{center}

\subsection{RQ2: What are Currently Unsatisfied Requirements}
\begin{figure}
    \centering
    \includegraphics[width=0.875\columnwidth]{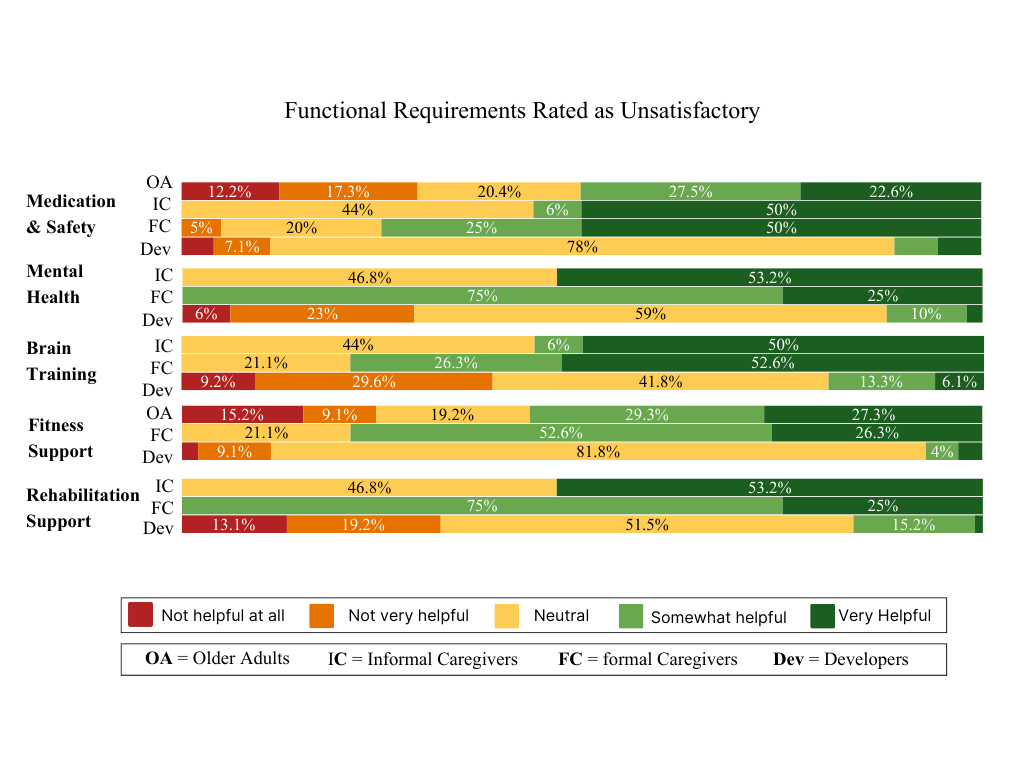}
    \caption{Functional requirements rated as unsatisfactory: Current satisfaction levels across stakeholder groups, showing proportion distributions.}
    \label{fig:cur_func_unsatis}
\end{figure}
\begin{figure}
    \centering
    \includegraphics[width=0.875\columnwidth]{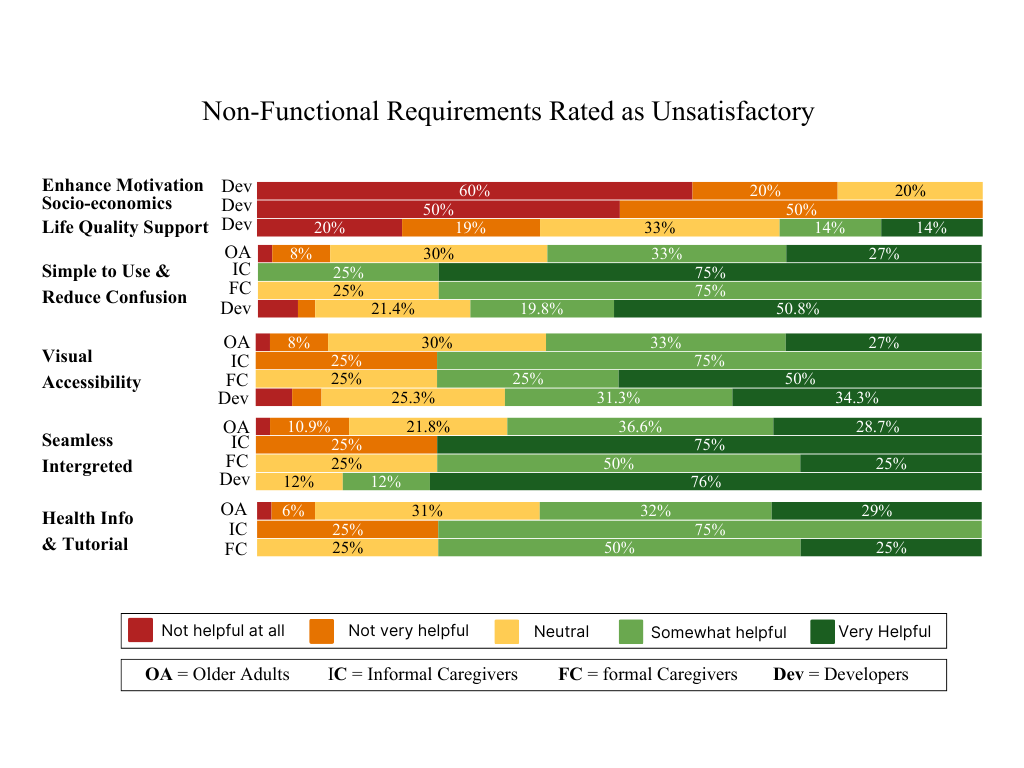}
    \caption{Non-Functional requirements rated as unsatisfactory: Current satisfaction levels across stakeholder groups, showing proportion distributions.}
    \label{fig:cur_nonfunc_unsatis}
\end{figure}

Figures~\ref{fig:cur_func_unsatis} and~\ref{fig:cur_nonfunc_unsatis} present the distribution of Likert-scale responses capturing stakeholders’ evaluations of currently unsatisfied functional and non-functional requirements. Each bar corresponds to a stakeholder group—older adult end users, caregivers (formal and informal), and developers—and indicates the percentage of responses across the five response categories, ranging from ``not helpful at all" to ``very helpful.'' These figures provide an overview of perceived shortcomings in current requirements and show how such perceptions vary across stakeholder groups.

\subsubsection{Unsatisfied Functional Requirements}
\paragraph{\textbf{Medication and Safety Support}}
Medication and safety support emerged as a widely recognised but inconsistently implemented requirement across stakeholder groups. As shown in Figure~\ref{fig:cur_func_satis}, while older adults and formal caregivers rated this area positively overall (means 4.4 and 4.3, respectively), developers were more neutral, with one-third selecting the midpoint category and only a minority rating it ``very helpful.” This divergence suggests that, although medication support is rated as satisfactory, its current design features may not fully address everyone's real-world complexity requirements. Qualitative feedback reinforces this mismatch: Programmers reported medication recognition and refill prompts as key future planned features. Older adults frequently mentioned difficulty managing multiple prescriptions or the lack of adherence when older patients use multiple medications. As OA96 described, ``I want a quick heads-up if my medicines or supplements don’t mix well—catching conflicts early keeps me safe.”  and FC18 mentioned, "Reminders help, but older adults sometimes cannot follow the correct steps...they need to follow medication limits, especially when they feel the medication is working well". These findings highlight a key improvement opportunity: medication safety tools must evolve from basic reminder functions toward integrated, adaptive systems that reflect diverse medication routines and literacy levels.
\paragraph{\textbf{Rehabilitation Support}}
Rehabilitation support provides recovery exercises tailored to rehabilitation requirements, which received mixed evaluations across stakeholder groups. Formal caregivers were positive, with 47\% reporting rehabilitation as ``neutral”. Developers were more reserved, with more than 82\% neutral and negative. 3 older adults noted it in an open-ended question, revealed that available apps often lack practical utility for recovery contexts. As OA76 commented on difficulties in following unclear guidance, with one saying, ``I couldn’t tell if I was doing the exercises correctly … the instructions were too vague…should focus on …safe and functional for recovery…don’t have to spend a lot to adapt their homes.” Dev27 commented, ``We need to make users feel guided through their exercises, not just watching a silent video.” Overall, these findings highlight a gap between caregivers’ optimism and users lived experience, suggesting that future rehabilitation tools must emphasize personalization, safety, and interactive feedback to support sustainable recovery rather than one-size-fits-all instruction.
\paragraph{\textbf{Mental Health Support}}
Although caregivers rated mental health support relatively positively, developers and older adults identified clear limitations. Nearly one-third of developers and a comparable proportion of older adults found current digital tools unhelpful, citing a lack of empathy and personal connection. Caregiver feedback described digital support as ``useful but impersonal,” while older adults emphasized that self-guided interventions rarely met their emotional needs. As OA19 reflected, ``When I feel anxious, I need a conversation, not just a checklist of relaxation tips.” Similarly, FC28 noted that mobile programs ``don’t make the elderly feel truly listened to.” These accounts reveal that while mental health features are valued conceptually, their current implementation lacks human warmth, responsive interaction, and personalized encouragement—qualities crucial for emotional engagement and trust.
\paragraph{\textbf{Fitness Support}}
Fitness support features were viewed as beneficial in principle but inconsistently effective in practice (Figure~\ref{fig:recruitment_workflow}). Caregivers appreciated their motivational potential, whereas developers and older adults reported limited usefulness or unclear instructions. Roughly one-third of all participants were neutral, suggesting that current fitness components neither frustrate nor fully engage users. Few older adults mentioned using such features regularly, describing them as repetitive or poorly adapted to individual capacity. OA29 noted, ``I’d like a smart, voice-interactive health assistant that connects with other devices. It could bring all the data together (in one place) so I can see my overall health."

\paragraph{\textbf{Brain Training}}
Brain training tools include memory games and cognitive exercises. Older adults showed mixed reactions, with 40\% neutral, 25\% ``somewhat helpful,” and 20\% ``very helpful”. Formal caregivers leaned less positive, with over a third neutral and 30\% finding it ``not helpful at all”. Developers were the most critical, with more than half rating it toward the unhelpful end, indicating dissatisfaction with current brain training tools (mean 2.8). Qualitative feedback revealed important shortcomings. OA48 noted, ``…I want a less invasive (training) tool…like a memory game…something that helps me take action before my memory gets worse.” These limitations highlight that while brain training is recognized as useful, current implementations are often too generic and fail to provide the depth and adaptability needed to support cognitive health effectively.

\paragraph{\textbf{Nutrition and Dietary Management}}
This requirement was newly identified from the qualitative data. Nutrition and dietary management were raised by 11 older adults as an important current function, though it was not frequently mentioned in future requirements, suggesting general satisfaction but limited expectations for further innovation.

\paragraph{\textbf{Pain Management}}
This requirement was also newly identified from the qualitative data. Pain management was referenced by 4 older adults as a future need, while no stakeholders emphasized it as a current feature, suggesting this area remains underdeveloped in existing Digital Health solutions and should be considered in future design efforts.

\begin{figure*}
  \centering
    \includegraphics[width=0.875\textwidth]{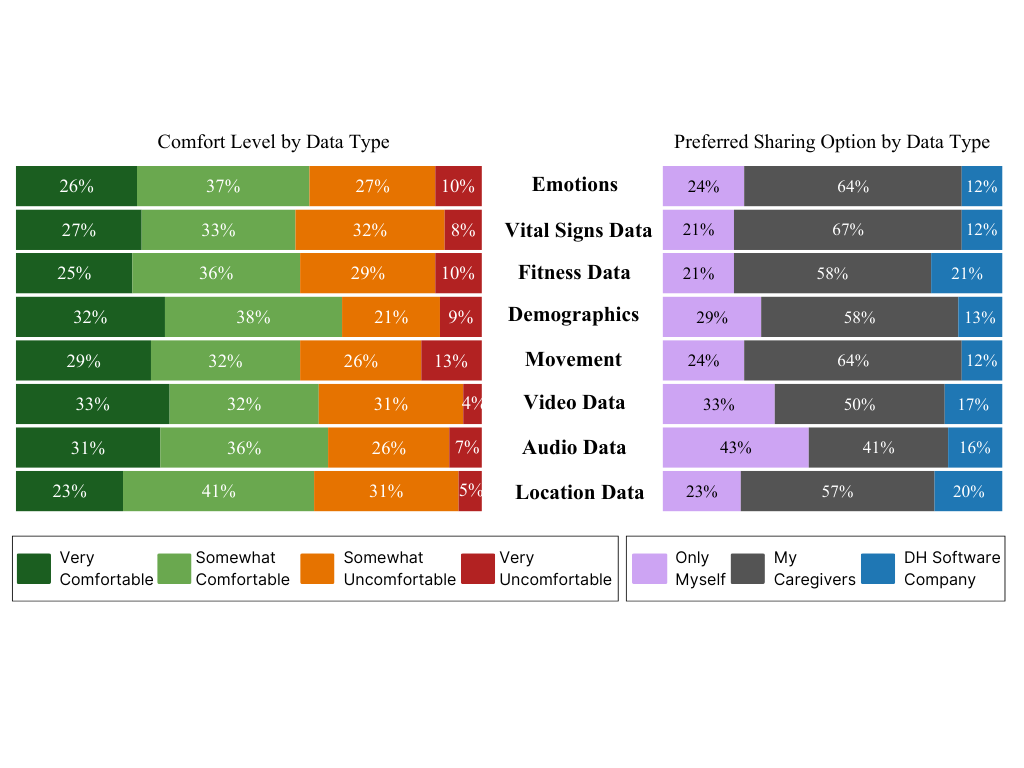}
    \caption{Data Collection Methods Trustworthy Reported by Older Adults: Who Older Adults want to share data with vs How Comfort Older Adults are to share}
    \label{fig:data_comfort}
\end{figure*}

%%\clearpage
\subsubsection{Unsatisfied Non-Functional Requirements}
\paragraph{\textbf{Socio-economic Support}}
Socio-economic support functions were rated lowest overall, reflecting a critical blind spot in design (Figure~\ref{fig:cur_func_unsatis}). Developers uniformly deprioritized affordability and device limitations, while older adults highlighted them as major barriers to engagement. As OA51 shared, ``My phone is old, and the screen is scratched — I can’t see clearly, but I can’t afford a new one.” Dev19 admitted, ``Affordability wasn’t part of our design brief; we focused on medical accuracy and safety.” These contrasting perspectives confirm that financial access remains an overlooked determinant of digital inclusion, underscoring the need for policy-level and design-level strategies that address affordability alongside functionality.

With respect to data governance (Figure~\ref{fig:data_comfort}), older adults were most willing to share health-related data with their caregivers rather than with software companies. Item-level analysis shows that security and privacy items received moderate current satisfaction ratings across groups (OA: M $\approx$ 3.6–3.8; Dev: M $\approx$ 3.9–4.1), with no statistically significant between-group differences at current time points. Qualitative responses revealed three dominant governance themes: conditional trust ('willing to share with caregivers but not companies'), transparency expectations ('want to see exactly who accessed my data'), and regulatory awareness differences between developers and end-users, with developers more likely to frame governance in terms of compliance while older adults framed it in terms of personal control.
\paragraph{\textbf{Life Quality Support}}
Life-quality support—features addressing wellbeing beyond medical tracking—received modest endorsement (Figure~\ref{fig:cur_func_unsatis}). Older adults described interest in tools that enhance rest, relaxation, and enjoyment, whereas developers maintained a narrower focus on clinical outcomes. ``I wish the app helped me sleep better or enjoy daily things, not just track my health,” OA88 explained. However, Developers have a prioritisation gap, like what Dev27 states, ``We plan to include wellbeing features later, but right now our focus is purely medical.” These findings illustrate a persistent mismatch between user expectations of holistic care and developers’ emphasis on functional health metrics, suggesting an opportunity to integrate wellbeing-oriented design for future development.
\paragraph{\textbf{Enhance Users’ Motivation}}
Motivational features, such as rewards or gamification, were the least supported across all groups (Figure~\ref{fig:cur_func_unsatis}). Developers often considered them unnecessary, while older adults viewed them as peripheral to genuine health management. OA52 noted, ``Reminders are enough for me — I don’t need points or rewards to look after my health.” Dev40 echoed this perspective, noting design trade-offs: ``Gamification sounds good, but it complicates design and doesn’t always fit health goals.” This highlights that motivational features are not yet widely perceived as necessary in health app contexts.

\paragraph{\textbf{Seamless Integration}}
Stakeholders uniformly recognised the importance of seamless integration but differed in how achievable it seemed (Figure~\ref{fig:cur_func_unsatis}). Older adults expressed frustration with fragmented systems, and caregivers highlighted the burden of managing multiple platforms. Developers agreed integration was critical but cited persistent technical barriers. As OA63 remarked, ``I don’t want five different apps; everything should be in one place so I don’t lose track.” Dev45 similarly commented, ``Integration with hospital systems is always the hardest part — everyone agrees it’s important, but it’s rarely achieved.” Together, these perspectives reveal that integration is a top user expectation for the future. However, due to the data privacy policies of different platforms, it can pose an ethical and regulatory concern rather than just a technical challenge.
\paragraph{\textbf{Simple to Use and Reduce Confusion}}
Simplicity was valued by all groups but inconsistently achieved in practice, as shown in Figure~\ref{fig:cur_nonfunc_satis}. Caregivers viewed clear navigation as essential, while developers and older adults diverged in satisfaction levels. Nearly one-third of developers rated current designs as only moderately helpful, and older adults’ mixed responses reflected ongoing confusion during use. Many participants described needing clearer icons and fewer steps. Qualitative accounts echoed these variations: Dev2 explained, ``We tried to reduce the number of steps, but balancing simplicity with function is harder than expected.” while OA44 noted, ``...fewer buttons and clearer words...(or I) forget what to press.” These results show that while simplicity is widely desired, older adults’ divided ratings and the spread among developers highlight that existing solutions still fall short of universal clarity and ease.

\paragraph{\textbf{Non-functional Requirements (data Governance)}}

As shown in Figure~\ref{fig:data_comfort}, older adults were most willing to share health-related and activity data with their caregivers rather than with software companies. Over half agreed to share health indicators, emotions, and movement data with caregivers ($\approx$ 63–66\%), while fewer than 20\% were comfortable sharing the same information with companies ($\approx$ 12–21\%). Comfort ratings reflected this caution: only about one-quarter of respondents reported feeling ``very comfortable,” whereas 30–40\% felt only ``somewhat comfortable,” and up to one-third expressed discomfort depending on the data type. This pattern suggests partial trust limited to caregiving contexts, rather than broad confidence in digital data sharing. OA summarized this concern: ``I want my partner to see my health tracks, but I need strong protection for my health info from strangers”.
\begin{center}
\begin{myframe}[\centering\textbf{RQ2 Answer Summary}]
\footnotesize
Collectively, these findings suggest that developers have largely been \textbf{\textit{``going for the low-hanging fruit}}. Current digital health systems are less successful in addressing requirements that are more \textit{``newly developed"} for digital health software, \textit{``highly specific to digital health"}, or \textit{``difficult to implement in practice"}. In particular, the currently unsatisfied functional requirements tend to involve care-intensive and higher-level support functions, which are in fact \textbf{more relevant to users with poorer health conditions}. Meanwhile, the unsatisfied non-functional requirements are more closely related to motivation, socio-economic support, and quality-of-life concerns, which are likewise more likely to matter for \textbf{users with lower self-efficacy or poorer health}. By contrast, more established usability-oriented qualities, such as simplicity, visual accessibility, and system integration, appear to be relatively better addressed. The lesson here is that technical maturity seems to favour requirements that are easier to standardise and implement, whereas \textbf{more human-centred, behavioural, and context-sensitive requirements} remain insufficiently supported in current digital health designs.

\end{myframe}
\end{center}

%\clearpage
\subsection{RQ3: What are Future Requirements for Innovation}
Future requirements for innovation were explored using two complementary types of questions. The first consisted of Likert-scale items that assessed how strongly participants desired specific requirements in future digital health systems. The second involved open-ended questions that allowed participants to elaborate on the functions and qualities they would like to see in future software.
\subsubsection{Future Functional Requirements}
\begin{figure}
  \centering
    \includegraphics[width=0.9\columnwidth]{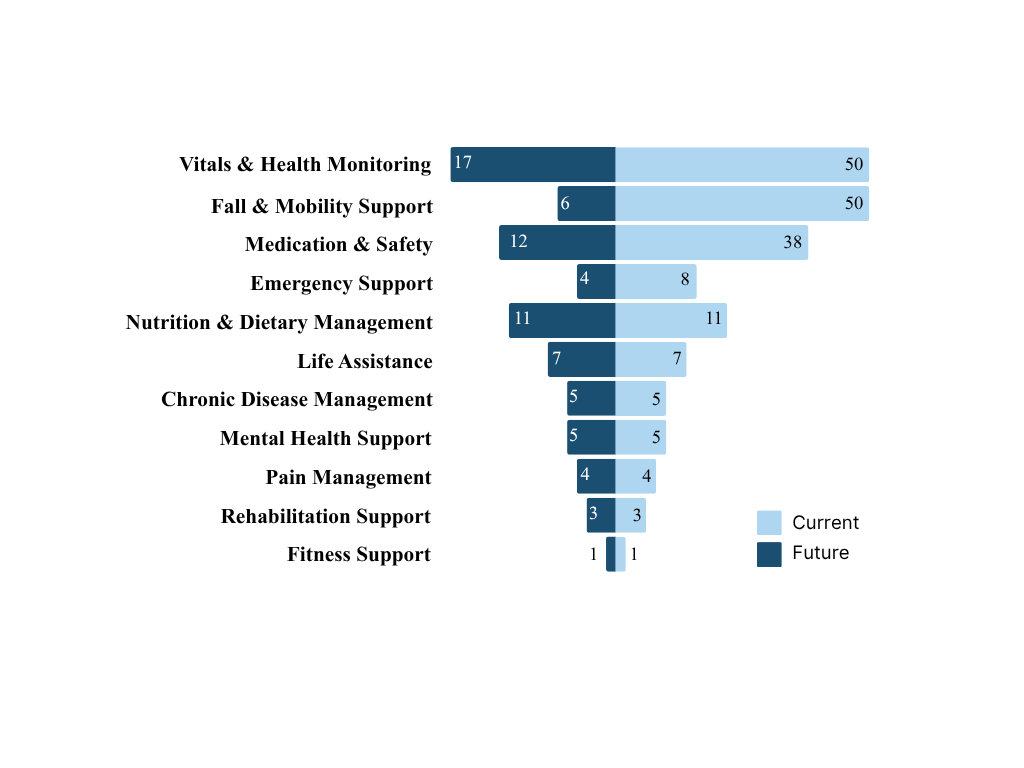}
    \caption{Frequency of Participants-Reported Functional Requirements in Open-ended Questions: Current Feedback vs Future Priority}
    \label{fig:freq_fun}
    \footnotesize
    \textit{Note.} These were general, non-leading questions asking participants to identify what they considered most important. The frequency counts reflect participants' primary areas of focus.
\end{figure}

As it summarised in Figure~\ref{fig:freq_fun}, for future design priorities, participants showed a distribution comparable to current importance ratings, with more counts for current importance than future priorities. ``Vital Signs and Health Monitoring” emerged as the most frequently mentioned function for both current and future perspectives, followed by ``Life Assistance” and ``Chronic Disease Management” (Figure~\ref{fig:cur_func_satis}). Both ``Life Assistance” and ``Chronic Disease Management” were mentioned a similar number of times—seven and five instances, respectively—across current importance and future suggestions. ``Fall and Mobility Support” was mentioned 50 times in relation to current feedback, but only 6 times as a future priority. When combined with the uneven satisfaction levels across participant groups—with older adults reporting dissatisfaction while caregivers and developers expressed higher satisfaction—this pattern (Figure~\ref{fig:cur_func_unsatis}), along with its relatively high future desirability ratings, suggests that older adults are dissatisfied with the current fall-support features and expect further improvement in future designs.

\subsubsection{Future Non-Functional Requirements}
\begin{figure}
  \centering
    \includegraphics[width=0.9\columnwidth]{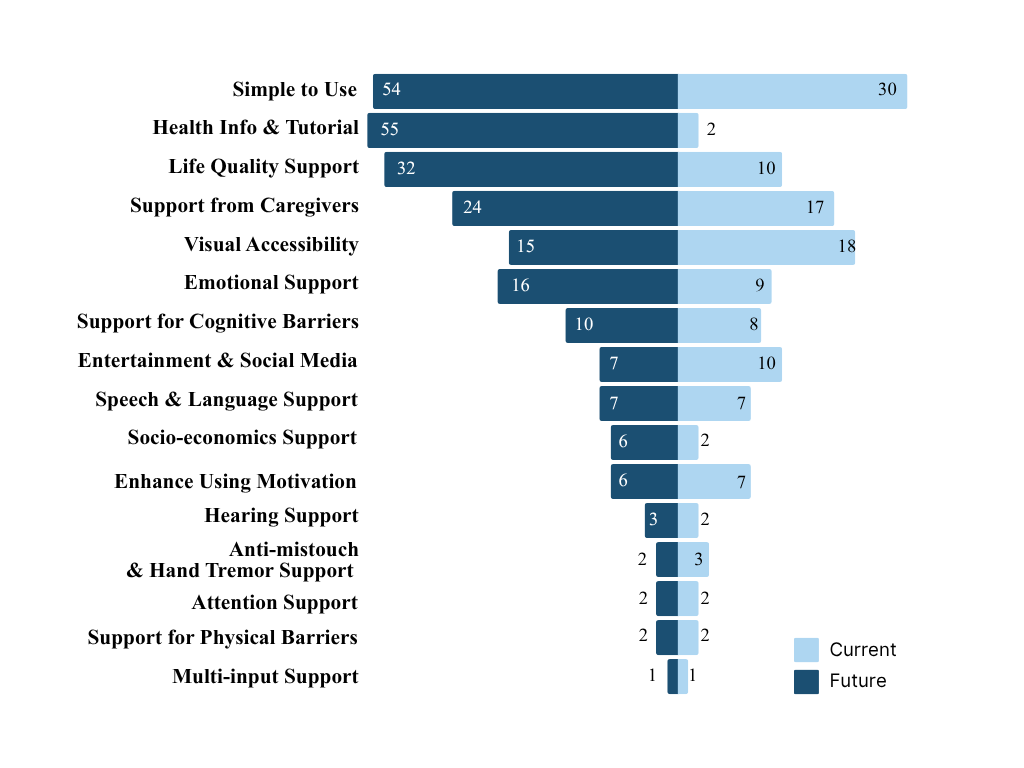}
    \caption{Current Satisfaction of Non-Functional Requirements among Stakeholder Groups: Proportion Distribution Stacked Bar Chart and Likert Scale Means (95\% CI)}
    \label{fig:nfun_cur_ci}
    \footnotesize
    \textit{Note.} These were general, non-leading questions asking participants to identify what they considered most important. The frequency counts reflect participants' primary areas of focus.
\end{figure}
In Figure ~\ref{fig:nfun_cur_ci}, visual accessibility received the highest satisfaction, particularly from older adults and caregivers, who rated larger fonts and clearer contrast as essential for maintaining independence. Developers, however, were less uniformly positive, revealing a continuing gap in aligning design emphasis with user accessibility expectations. As Dev8 stated, ``We designed the interface to conform to larger fonts and colour contrast standards,” and OA12 explained, ``I only like black text on my mobile; it should be bold as well. The colourful title does not stand out for me.”

Key future non-functional innovations requirements are what participants desired: personalisation, health information and tutorials, quick responsiveness, seamless integration, simple-to-use design, and visual accessibility. The full percentage proportion data and mean can be found in the Multimedia Appendix. 

Some requirements—such as personalisation and health information/tutorials—are already valued and relatively well implemented, while simplicity, responsiveness, and integration were previously discussed as current limitations (i.e., important but not consistently effective).

The first direction relates to usability and workflow improvements—simplicity of use, quick responsiveness, visual accessibility, and seamless integration—where developers expressed higher satisfaction (means $\approx$ 3.8) than older adults (means $\approx$ 3.7) and caregivers (means $\approx$ 4.5). Over 60\% of developers considered these functions already sufficient, yet around one-third of older adults remained neutral or dissatisfied, particularly regarding slow responses and fragmented records. These gaps confirm that technical optimisation alone does not ensure usability or trust, emphasizing the need for future systems to prioritise reliability, speed, and accessibility that align with senior users’ expectations rather than developer assumptions.

The second direction focuses on personalised accessibility, encompassing personalisation and health information/tutorials. Older adults and caregivers showed the strongest enthusiasm, with more than 70\% of older adults and 75\% of caregivers rating these features as ``very much desired” (means $\approx$ 4.7), compared with developers’ more moderate interest (means $\approx$ 3.9). Furthermore, ``Health information and tutorials” was mentioned as the second most frequent category in open-ended responses, highlighting its importance. Collectively, these findings indicate that even features currently perceived as satisfactory could benefit from further development, possibly reflecting older adults’ low expectations for personalised accessibility and the opportunity to exceed them through adaptive, user-centred innovation.

Unlike the functional requirements, the frequency of participants mentioning future non-functional requirements was considerably higher than the frequency of referencing current features. Most of the non-functional requirements showed a similar distribution pattern between future and current perspectives. However, ``Health Information and Tutorial” stood out — it was mentioned only twice in current feedback but 55 times when participants discussed future priorities, making it the most frequently cited future need.
When combined with its low satisfaction rating in Figure~\ref{fig:cur_nonfunc_unsatis}, this finding suggests a clear demand for improvement and redesign of this feature in future implementations, particularly to better meet the expectations of older adults and caregivers. This need appears especially salient for formal caregivers such as clinicians, who often rely on accessible educational tools to support patient understanding. As FC30 noted, ``Using apps to help them (older adults) get admitted and discharged more conveniently and timely in the hospital, get their attention and explain the situation for the older adults.” Similarly, OA78 highlighted, ``All of my friends are using … App; I also use it to communicate with my kids, sharing wellness information with them every day.” Together, these comments underline that improving the ``Health Information and Tutorial” function, particularly by integrating it with ``Support from Caregivers,” could substantially enhance usability and perceived value for older adult users.

Furthermore, in relation to ``Visual Accessibility,” future priorities shifted toward speech, hearing, and cognitive support, which were mentioned more frequently by older adults than by developers or caregivers. This discrepancy highlights that while current systems perform well in visual design, other accessibility domains—especially those addressing sensory or cognitive barriers—remain underdeveloped. As OA58 observed, ``I think multiple input methods are good, so I can just talk to it if my finger is not working that day. I cannot type fast, and my mom cannot type at all.”
\begin{center}
\begin{myframe}[\centering\textbf{RQ3 Answer Summary}]
\footnotesize
Future functional design priorities largely \textit{mirrored the currently satisfied patterns}, with ``vital signs and health monitoring" remaining the most prominent requirement, followed by ``life assistance" and ``chronic disease management". At the same time, future non-functional design priorities were more strongly \textit{aligned with the currently unsatisfactory} requirements, highlighting a clear need to improve and redesign ``health information and tutorial features", particularly to better support both caregivers and older adults through accessible educational tools; as FC30 noted, future systems should provide a tool that can ``\textbf{\textit{seize the attention and bridge their understanding}} of the situation."

\end{myframe}
\end{center}

%\clearpage
\subsection{RQ4. What are the Stakeholder Perception Gaps}
\begin{table*}[t]
\centering
\scriptsize
\setlength{\tabcolsep}{3pt}
\renewcommand{\arraystretch}{1.15}
\begin{threeparttable}
\caption{One-Way ANOVA Results for Perceived Importance of Functional Requirements Across Stakeholder Groups: Current vs.\ Future}
\label{tab:anova_functional_merged}
\begin{tabular}{lp{1.2cm}p{0.7cm}p{0.7cm}p{0.7cm}p{0.7cm}p{1cm}p{1cm}p{1cm}c}
\toprule
\textbf{Requirements}
  & \textbf{Type}\tnote{d}
  & \textbf{OA}\tnote{b}
  & \textbf{Dev}\tnote{b}
  & \textbf{IC}\tnote{b}
  & \textbf{FC}\tnote{b}
  & \textbf{F}
  & \textbf{$p$}
  & \textbf{$\eta^2$}
  & \textbf{Perception Order}\tnote{c} \\
\midrule

% %--- Satisfactory ---
% \multicolumn{10}{l}{\textit{Satisfactory functional requirements (RQ1.1 / RQ4.1)}} \\[2pt]

Chronic Disease \ Mgmt
  & Current & 3.7 & 3.0 & 5.0 & 4.0
  & 13.572  & $<$.001 & 0.199
  & Dev $\ll$ OA $<$ FC $\approx$ IC \\
  & Future  & 4.0 & 3.7 & 4.4 & 2.0
  & 7.115  & $<$.001 & 0.085
  & Dev $\approx$ OA $<$ FC $\approx$ IC \\[4pt]
Fall \& Mobility Supp.
  & Current & 3.3 & 3.0 & 4.5 & 4.8
  & 9.666  & $<$.001 & 0.150
  & Dev $\ll$ OA $<$ FC $\approx$ IC \\
  & Future  & 3.9 & 3.8 & 4.4 & 4.8
  & 9.23  & $<$.001 & 0.180
  & Dev $<$ OA $<$ IC $<$ FC \\[4pt]
Life Assistance
  & Current & 3.0 & 4.5 & 4.5 &3.0
  & 7.642  & $<$.001 & 0.123
  & FC $\approx$ OA $<$ Dev $<$ IC \\
  & Future  & 3.9 & 3.8 & 4.3 & 4.7
  & 7.118  & $<$.001 & 0.085
  & Dev $\approx$ OA $<$ FC $\approx$ IC \\[6pt]
Medication \& Safety
  & Current & 4.4 & 3.0 & 4.0 & 4.3
  & 4.304  & .006    & 0.070
  & Dev $<$ OA $\approx$ FC $\approx$ IC \\
  & Future  & 4.0 & 3.9 & 4.6 & 3.0
  & 6.734  & $<$.001 & 0.080
  & Dev $\approx$ OA $<$ FC $\approx$ IC \\[4pt]
Mental Health\tnote{a}
  & Current & ---              & 3.0 & 4.0 & 4.3
  & 229.6  & $<$.001 & 0.818
  & Dev $\ll$ FC $\approx$ IC \\
  & Future  & \multicolumn{7}{l}{\textit{---}} \\[4pt]
Brain Training
  & Current & 3.5 & 2.8 &  4.0 & 4.3
  & 9.640  & $<$.001 & 0.150
  & Dev $\ll$ OA $\approx$ FC $\approx$ IC \\
  & Future  & 3.9 & 3.9 & 4.7 & 4.6
  & 6.873  & $<$.001 & 0.082
  & Dev $\approx$ OA $<$ FC $\approx$ IC \\[4pt]
Fitness Support\tnote{a}
  & Current & 3.6 & 3.0 & 4.0 & 5.0
  & 18.600 & $<$.001 & 0.254
  & Dev $<$ OA $\approx$ FC $\approx$ IC \\
  & Future  & \multicolumn{7}{l}{\textit{---}} \\[4pt]
Rehabilitation Supp.
  & Current & 2.7 & 4.0 & 4.3 &2.7
  & 15.420 & $<$.001 & 0.220
  & Dev $<$ OA $\approx$ FC $\approx$ IC \\
  & Future  & 3.6 & 3.8 & 4.7 & 3.0
  & 8.229  & $<$.001 & 0.096
  & Dev $\approx$ OA $<$ FC $\approx$ IC \\

\bottomrule
\end{tabular}
\begin{tablenotes}[flushleft]
\footnotesize
\item$F$ is the one-way ANOVA test statistic, computed as the ratio of between-group variance to within-group variance. Larger $F$ values indicate that differences among group means are larger relative to variability within groups.
\item$p$ is the significance level associated with the ANOVA test. A small $p$ value (typically $p<.05$) indicates that the observed group differences are unlikely to be due to random variation alone, suggesting a statistically significant difference among the stakeholder groups.
\item$\eta^2$ was calculated as $SS_{\text{between}} / SS_{\text{total}}$. SS: Sum of Squares of score in Anova test.
\item[a] --- means do not have enough answers to be sampled.
\item[b] OA = Older Adults; Dev = Developers; IC = Informal Caregivers; FC = Formal Caregivers. Mean value.
\item[c] \textit{Perception Order} reflects the ordering of group means based on post-hoc pairwise comparisons. Tukey HSD was applied when Levene's test indicated homogeneity of variance; Games-Howell was used when variances were unequal. '<<' denotes a highly significant difference (p < .001); '<' denotes a significant difference (p < .05). Rankings are intended to highlight perception gaps between stakeholder groups and inform design priorities for developers.
\item[d] Type of perception. Current = stakeholders' perceived satisfactory levels of existing systems; Future = stakeholders' desired levels for future systems.
\end{tablenotes}
\end{threeparttable}
\end{table*}

\begin{table*}[!ht]
\centering
\scriptsize
\setlength{\tabcolsep}{3pt}
\renewcommand{\arraystretch}{1.15}
\begin{threeparttable}
\caption{One-Way ANOVA Results for Non-Functional Requirements Across Stakeholder Groups: Current vs.\ Future}
\label{tab:anova_nonfunc_merged}
\begin{tabular}{lp{1cm}p{0.7cm}p{0.7cm}p{0.7cm}p{0.7cm}p{1cm}p{1cm}p{1cm}c}
\toprule
\textbf{Requirements}
  & \textbf{Type}\tnote{d}
  & \textbf{OA}\tnote{b}
  & \textbf{Dev}\tnote{b}
  & \textbf{IC}\tnote{b}
  & \textbf{FC}\tnote{b}
  & \textbf{F}
  & \textbf{$p$}
  & \textbf{$\eta^2$}
  & \textbf{Perception Order}\tnote{c} \\
\midrule

% %--- Satisfactory ---
% \multicolumn{10}{l}{\textit{Satisfactory non-functional requirements (RQ1.2 / RQ4.1)}} \\[2pt]

Entertainment/Social
  & Current & 3.7 & 4.0 & 4.7 & 4.0
  & 0.29 & .750 & .008
  & OA $\approx$ Dev $\approx$ FC $\approx$ IC \\
  & Future  & 3.9 & 4.0 & 3.5 & 4.6
  & 23.455& $<$.001&0.213&IC $<$ OA $\approx$Dev  $<$ FC \\
Personalisation
  & Current & 4.0 & 4.0 & 4.1 & 4.0
  & 0.16 & .853 & .005
  & OA $\approx$ Dev $\approx$ FC $\approx$ IC \\
  & Future  & 4.1 & 3.8 & 4.5 & 4.3
  & 4.48 & .012 & .040
  & OA $\approx$ FC $<$ IC \\[4pt]
Quickly Responsive\tnote{a}
  & Current & 3.9 & ---              & 4.7 & 4.3
  & 2.13 & .127 & .059
  & OA $\approx$ FC $\approx$ IC \\
  & Future  & 3.8 & 3.8 & 5.0 & 4.6
  & 33.22 & $<$.001 & .278
  & OA $<$ FC $<$ IC \\[6pt]
Simple to Use
  & Current & 3.8 & 3.9 & 4.8 & 3.5
  & 1.29 & .278 & .012
  & OA $\approx$ Dev $\approx$ FC $\approx$ IC \\
  & Future  & 4.0 & 3.9 & 4.0 & 4.7
  & 39.68 & $<$.001 & .185
  & OA $<$ FC $\approx$ IC \\[4pt]
Visual Accessibility\tnote{a}
  & Current & 3.8 & 3.9 & 3.5 & 4.3
  & 1.17 & .320 & .164
  & FC $\approx$ IC \\
  & Future  & ---              & 3.9 & 5.0 & 4.7
  & 23.61 & $<$.001 & .214
  & OA $<$ FC $<$ IC \\[4pt]
Seamless Integrated
  & Current & 3.7 & 4.6 & 4.6 & 4.3
  & 0.94 & .394 & .027
  & OA $\approx$ Dev $\approx$ FC $\approx$ IC \\
  & Future  & 4.0 & 3.8 & 5.0 & 4.4
  & 10.09 & $<$.001 & .104
  & OA $<$ FC $<$ IC \\[4pt]
Health Info \& Tutorial\tnote{a}
  & Current & 3.8 & ---              & 4.7 & 4.0
  & 0.51 & .600 & .015
  & OA $\approx$ FC $\approx$ IC \\
  & Future  & 4.0 & 3.8 & 5.0 & 4.6
  & 23.455& $<$.001&.213  & FC $<$ OA $\ll$ IC\\
\bottomrule
\end{tabular}
\begin{tablenotes}[flushleft]
\footnotesize
\item$F$ is the one-way ANOVA test statistic, computed as the ratio of between-group variance to within-group variance. Larger $F$ values indicate that differences among group means are larger relative to variability within groups.
\item$p$ is the significance level associated with the ANOVA test. A small $p$ value (typically $p<.05$) indicates that the observed group differences are unlikely to be due to random variation alone, suggesting a statistically significant difference among the stakeholder groups.
\item$\eta^2$ was calculated as $SS_{\text{between}} / SS_{\text{total}}$. SS: Sum of Squares of score in Anova test.
\item[a] --- means do not have enough answers to be sampled.
\item[b] OA = Older Adults; Dev = Developers; IC = Informal Caregivers; FC = Formal Caregivers. Mean value.
\item[c] \textit{Perception Order} reflects the ordering of group means based on post-hoc pairwise comparisons. Tukey HSD was applied when Levene's test indicated homogeneity of variance; Games-Howell was used when variances were unequal. '<<' denotes a highly significant difference (p < .001); '<' denotes a significant difference (p < .05). Rankings are intended to highlight perception gaps between stakeholder groups and inform design priorities for developers.
\item[d] Type of perception. Current = stakeholders' perceived satisfactory levels of existing systems; Future = stakeholders' desired levels for future systems. 
\end{tablenotes}
\end{threeparttable}
\end{table*}
Tables~\ref{tab:anova_functional_merged} and \ref{tab:anova_nonfunc_merged} compare perception gaps across stakeholder groups for functional and non-functional requirements at current and future time points. Overall, the results show that these gaps vary by requirement type and temporal focus. Functional requirements exhibit more consistent differences across groups in both current evaluations and future expectations, whereas non-functional requirements show relatively limited divergence in current perceptions but clearer separation in future-oriented ratings. Across both domains, caregivers, particularly informal caregivers, tend to assign higher future ratings than older adults, indicating that expectations for future systems are shaped not only by system use, but also by caregiving responsibilities and practice contexts.
\subsubsection{Perception Gaps in functional Requirements}
Table~\ref{tab:anova_functional_merged} presents one-way ANOVA results for all functional requirement items across stakeholder groups, comparing current and future perceptions side by side. All eight requirements show statistically significant between-group differences in both time points (all $F > 4.30$, all $p \leq .006$), confirming that perception gaps are a persistent rather than incidental feature of the stakeholder landscape. 
 
\paragraph{\textbf{Developers as a consistent outlier.}}
The most consistent pattern across current perception ratings is that Developers
occupied the lowest position in the perception order for every item. This was most pronounced for Mental Health, where the gap between Developers and caregiver groups was statistically the largest in the study, followed by Fitness Support ($\eta^2 = .254$), Rehabilitation Support ($\eta^2 = .220$), and Chronic Disease Management ($\eta^2 = .199$). Post-hoc comparisons revealed three distinct current patterns: a full ordered chain where Developers rated significantly lower than
Older Adults, who in turn rated lower than caregivers (Chronic Disease Management, Fall \& Mobility Support, Fitness Support); Developers as the sole outlier below a statistically equivalent non-developer cluster (Brain Training, Rehabilitation Support, Medication \& Safety); and a caregiver-versus-non-caregiver binary split where Developers and Older Adults clustered together below both caregiver groups (Life Assistance, Mental Health). In all cases, Formal and Informal Caregivers did not differ significantly from each other ($p \geq.245$ across all items), indicating that the primary differences occur between developers and all other stakeholders, instead of between caregivers and older adults.
 
\paragraph{\textbf{Convergence and persistent gaps for future innovation}}
It is worth noting that when comparing current and future perception orders, across all items for which future data were available, the gap between Developers and Older Adults narrows substantially: the future perception order consistently shows Dev $\approx$ OA $<$ \{FC $\approx$ IC\}, whereas the current order frequently showed Dev $<$ OA or Dev $\ll$ OA.
This suggests that developers' future aspirations converge toward those of older adults, yet neither group closes the gap with caregivers. For example, ``Brain Training" shifted from
Dev $<$ \{OA $\approx$ FC $\approx$ IC\} (current) to Dev $\approx$ OA $<$ \{FC $\approx$ IC\} (future),
and ``Rehabilitation Support" shifted from
Dev $<$ \{OA $\approx$ FC $\approx$ IC\}
to Dev $\approx$ OA $<$ \{FC $\approx$ IC\}.
Effect sizes also moderated between current and future for
several items --- for instance, ``Chronic Disease Management"
decreased from $\eta^2 = .199$ to $\eta^2 = .085$ ---
reflecting that the overall disagreement among all four groups
was smaller in future ratings, even though the caregiver--non-caregiver divide remains statistically significant.
Furthermore, ``Medication \& Safety" is particularly noteworthy. currently, it is the item with the smallest effect ($\eta^2 = .070$,
with Dev $<$ OA as the sole significant pair),
but in future ratings it shifts to Dev $\approx$ OA $<$ \{FC $\approx$ IC\}
($\eta^2 = .080$), indicating that developers and older adults now agree on a \textit{high} future priority for this item.
Caregiver groups also continue to rate this medication safety support near the ceiling of the scale (highest).
Taken together, these patterns suggest that while developer--end-user (and indirect-user) alignment improves modestly over the future horizon, the structural gap between developer and non-developer stakeholders is unlikely to be resolved without targeted participatory design interventions that embed caregiver and end users' engagement directly into requirements prioritisation.
\subsubsection{Perception Gaps in Non-Functional Requirements}
Table~\ref{tab:anova_nonfunc_merged} reports one-way ANOVA results for stakeholders' perceptions of satisfactory levels of non-functional requirements at current time points and for future desired innovation. What stands out in the table is that the current perceptions between older adults and caregivers did not show significant differences, whereas those between developers and end users with available data were greater but still not statistically significant (all $p \geq .127$). This, together with the qualitative data, suggests that caregivers and older adults have more similar perceptions in their evaluations of current non-functional qualities, and that these are lower than those of developers. The largest descriptive spread appeared in ``Seamless Integration'', where Dev reported the highest mean ($M = 4.6$) and OA the lowest ($M = 3.7$). Overall, perceptions of current non-functional requirements were comparatively uniform.
In contrast, future ratings revealed clear stakeholder differences, with Informal Caregivers (IC) emerging as the highest-rated group across all significant items. The largest effects were observed for Quickly Responsive ($\eta^2 = .278$), Visual Accessibility ($\eta^2 = .214$), and Seamless Integration ($\eta^2 = .104$), where ratings followed a consistent ordering of OA $<$ FC $<$ IC. This pattern suggests that expectations for future non-functional quality increase with caregiving involvement, with informal caregivers expressing the strongest demands.
A similar pattern appeared for ``Simple to Use'', although here the structure was a two-cluster split: OA and caregivers rated the item significantly lower than developer groups (OA $<$ \{FC $\approx$ IC\}; $\eta^2 = .185$, $F(2,349) = 39.68$, $p < .001$). This finding indicates that caregivers place greater emphasis on future usability improvements than older adults. Personalisation showed a more differentiated pattern ($\eta^2 = .040$): OA and FC did not differ significantly ($p = .106$), whereas IC rated the item significantly higher than both OA ($p < .001$) and FC ($p = .031$). This suggests that informal caregivers place particular value on systems that can be tailored to individual care needs.

\paragraph{\textbf{Intermediate position of formal caregivers between developers and older adults}}
Formal Caregivers (FC) generally occupied an intermediate position: their current ratings were broadly aligned with the other groups, whereas in future ratings they tended to rate items above OA and below, or similar to, IC. This pattern suggests that FC represent a middle position between end-user and informal caregiver expectations. Developer data were not available for three non-functional items, which limits direct comparison between developer and user expectations in this domain. This absence should be acknowledged as a limitation of the requirements elicitation process.
\paragraph{\textbf{Synthesis of functional and non-functional requirements}}
The perception gaps of functional and non-functional requirements reveal that these dimensions are deeply interdependent; the success of a core feature is often predicated on its underlying quality attributes. For example, while the functional requirement for 'medication management' was highly prioritised by all stakeholders, its real-world utility is compromised if non-functional requirements such as 'simplicity' and 'visual accessibility' are not fulfilled. As OA90 noted: 'An interface mode specific to older adults should not be complex and distracting. Unnecessary chatting features make me forget where the key functions I need are—like medication reminders.' This interaction necessitates a \textbf{\textit{co-design approach where stakeholders concurrently define system functions and behaviours}}. By addressing both dimensions simultaneously, developers can prevent creating technically accurate but practically unusable digital health solutions.
\begin{center}
\begin{myframe}[\centering\textbf{RQ4 Answer Summary}]
\footnotesize
In summary, three patterns characterise stakeholder perception
gaps. First, \textbf{\textit{Developers are a consistent outlier}} for almost all requirements --- suggesting that developer
undervaluation of senior end users' requirements in health-support features is a present-state problem. Second, Formal and Informal
Caregivers form a persistent joint caregiver cluster above both Developers and Older Adults across almost all requirements, whether current, future, functional, or nonfunctional. One interesting exception is ``Personalisation'', where IC rates significantly higher than FC, reflecting the individual-level customisation demands of \textbf{\textit{one-to-one informal care in contrast to the formal caregivers' multiple-client services}}. Third, non-functional requirements show large future divergence, suggesting that greater design attention should be directed to this area.\\
Overall, these patterns indicate that perception gaps are shaped by stakeholder role: current assessments reflect differences in existing experience, while future expectations reveal stronger divergence in anticipated care needs. This suggests that future design should not treat any single stakeholder group as representative of all others; \textbf{\textit{engaging stakeholders}}
is important for fully understanding the requirements of older adults, and \textbf{\textit{formal caregivers can help bridge the gap in clinical and health-related requirements}}, while \textbf{\textit{informal caregivers can help shape understanding of the technology proficiency gap}} in future design.
\end{myframe}
\end{center}
%\clearpage
\section{Recommendations for Research and Practice}\label{discussion-recommendations-for-research-and-practice}

% This study presents the first empirical three-group comparison of requirements perceptions across older adults, caregivers, and developers in aged-care digital health, revealing a systematic and role-structured Requirements Gap. Developers consistently undervalue older adults' NFR dissatisfaction while overestimating the adequacy of advanced functional features. A validated and extended requirements catalogue is produced, incorporating new items surfaced from open-ended responses. Together, these findings offer actionable stakeholder-gap evidence to guide co-design, product prioritisation, and privacy-by-design decisions in aged-care software development.

\textbf{Prioritise support for key functional areas:} Overall, satisfaction with current functional requirements was rated higher than non-functional ones, and older adults and their caregivers generally reported greater satisfaction than developers. This pattern aligns with recent evidence in two ways. First, it shows that the key motivation of older adults who use digital health software lies more in features that strengthen independence and quality of life and less in gamification and entertainment, which is opposite to what developers assume it to be. This aligns with findings from White et al and Wu et al, who found that patients, including older adults, who use gamification to enhance health software adherence care more about their health outcomes being improved by software, not the gamification itself~\cite{white_gamification_2022, chen_digital_2023}

Second, this difference reflects users’ higher sensitivity to reliability and responsiveness compared with developers’ emphasis on novelty, confirming that older end users behave similarly to other patient groups described in previous research on wearable trackers’ accuracy review ~\cite{singh_real-world_2024}. Developers should therefore consolidate progress in functional areas where users already express satisfaction. Functional domains such as medication management, fall prevention, and vitals monitoring were consistently endorsed by end users, aimed at making these functions more reliable and safe to use, a pattern comparable to findings by both Niyomyart et al and Portz et al ~\cite{niyomyart_current_2024,portz_using_2019}. For example, for ensuring adherence, managing complex drug interactions, and minimising side effects, future research should therefore co-design and validate medication modules with clinicians and pharmacists, integrating automated drug–drug interaction checks and contextual safety alerts to enhance trust and usability~\cite{singh_real-world_2024, wu_acceptability_2024}.

~

\textbf{Prioritise Senior User and Caregiver desired features:} The functional priorities identified by developers differed from those of older adults, with caregivers’ perspectives falling in between. This can be because caregivers can empathise with older adults more deeply and better understand their requirements. This phenomenon highlights the unique relational dynamic between senior users and their caregivers, where caregiving roles shape technology expectations and emotional engagement. This also accords with our earlier observations, which showed that patients’ technology adoption rate is higher when their caregivers use technology~\cite{lee_steps_2017,van_damme_perspectives_2020}. For example, caregivers ranked features such as emotional support and fitness as important, though they assigned slightly lower importance than older adults themselves. But it shows that both these groups acknowledge these features as important for continuity of care and preventive aspects of well-being, which was very different from developers who did not highlight these aspects. This finding is consistent with evidence from Zainal et al showing that digital health software can reduce caregiver burden while simultaneously improving the overall health outcomes and emotional well-being of care recipients~\cite{yang_effects_2024,zainal_exploring_2025}. These results reinforce that digital health design must recognise the interdependence between senior users and caregivers, focusing on features that sustain both usability and emotional priority across this dyad.

Besides, the perception differences observed between formal and informal caregivers suggest that these two groups should not be treated as a homogeneous caregiver population in future research. Future studies should ensure balanced and sufficient recruitment of both formal and informal caregivers as distinct groups, enabling more robust subgroup comparisons and a clearer understanding of how caregiving roles shape digital health technology requirements.

\textbf{Respect older adults' wisdom:} In contrast to previous studies~\cite{portz_using_2019,banbury_multi-site_2014}, we found that older adults were more discerning and critical than many developers assumed, particularly regarding non-functional requirements such as usability, accessibility, and system responsiveness. This difference suggests that developers and caregivers often generalise from the types of older patients they encounter most frequently and assume that older adults are less capable of using new technologies, which are understandable opinions but not fully representative. These results further support the idea that stereotypes of ageism exist in digital health design for older adults~\cite{mannheim_ageism_2023,mariano_too_2022}. While this is partly accurate—many do need additional accessibility support—our findings indicate that such assumptions are incomplete. A substantial proportion of older adults clearly understand their own health needs and show genuine enthusiasm for learning and using digital health software. Therefore, future research should take serious account of older adults’ opinions and respect their wisdom as end users with extensive life experience. 

\textbf{Address gaps in key non-functional areas:} Older adults were not fully satisfied with basic non-functional features that developers considered already solved, including simplicity of use, quick responsiveness, and seamless integration across systems. While developers expressed confidence in these areas, older adults provided more cautious evaluations, indicating that technical functionality does not necessarily translate into practical trust or satisfaction. This observation is consistent with qualitative evidence showing that excessive alerts, slow interfaces, and fragmented workflows can erode confidence even when tools perform as intended~\cite{mannheim_ageism_2023,mariano_too_2022,bertolazzi_barriers_2024,gani_understanding_2025,yang_effects_2024}. This disparity may be explained by the technological gap between developers and end users. Factors such as the types of sensors used for data collection, interface response speed, and the availability of tutorial information all influence system acceptance among both clinicians and end users. Future research should therefore establish standardised usability and interoperability benchmarks for older adults, including time-to-task metrics, error rates, and perceived ease-of-use indicators, to enable consistent comparison across studies and product types.

\textbf{Prioritised accessibility:} Compared with developers, caregivers—particularly formal caregivers—demonstrated greater awareness of the health factors influencing older adults’ technology adoption. This difference may stem from caregivers’ deeper understanding of older adults’ health conditions or their closer proximity to end users along the developer–user spectrum. These observations are consistent with recent evidence that barriers for older adults include usability hurdles as well as structural issues such as digital literacy and connectivity, which directly affect engagement with mHealth tools~\cite{niyomyart_current_2024,yang_effects_2024}. For future design, accessibility cannot be restricted to interface design alone; for instance, solving vision problems cannot rely merely on enlarging text size but on the holistic design of contrast, an elegant interface, and effortless interaction.

\textbf{Address wider human aspects:} Addressing human aspects like emotional needs is associated with better engagement among older adults in Digital Health software. This is consistent with previous caregiver-focused evaluations~\cite{huang_understanding_2025, bertolazzi_barriers_2024,an_older_2024}. Our findings further highlight the importance of socio-economic status, emotional well-being, and the degree of support available through both formal and informal caregiving networks. Hence, designers must move beyond a narrow technical view of accessibility and consider the wider human aspects that caregivers emphasise when shaping future co-design strategies. Together, these insights underscore the need for a holistic, empathy-driven design framework that values caregivers’ emotional understanding as a determinant of sustained engagement among older adults.

\textbf{Address cost issues in app usage:} Future designs should explicitly address cost-related issues. This finding is relatively novel yet broadly consistent with previous studies linking financial affordability with patient engagement~\cite{niyomyart_current_2024,yu_study_2025, an_older_2024,chadwick_engagement_2024}. Socio-economic conditions, the extent of access to or affordability of extra support from formal caregivers, and the cost of maintaining emotional well-being were named as significant game-changers by older adults as part of their requirements. While developers largely dismissed socio-economic support as outside the design scope, older adults’ comments about struggling with outdated or damaged devices underscore that financial and material conditions cannot be ignored. This aligns with earlier research demonstrating the relationship between patient health outcomes and financial disparities~\cite{chadwick_engagement_2024,hepburn_barriers_nodate,yang_digital_2024}. Syntheses of older-adult mHealth use identify affordability, device adequacy, and internet connectivity as critical determinants of ongoing participation~\cite{niyomyart_current_2024}. These findings emphasise that digital inclusion requires not only technical usability but also economic feasibility. 

\textbf{Adopt privacy-by-design as a first-class requirement.} Governance findings indicate that older adults extend only conditional trust to digital health systems, preferring caregiver-mediated data sharing over direct company access. Future designs should implement explicit consent interfaces that clearly distinguish caregiver-facing from company-facing data flows, granular access controls allowing older adults to specify sharing preferences per data type, and audit-trail features that make data access visible and interpretable to non-technical users. These features should be treated as core NFRs rather than post-hoc additions.

\textbf{Motivation and life quality support:} Caregivers identified this as important, but developers have overlooked the importance of life quality support in mHealth solutions. Although not universally prioritised, these are linked to emotional resilience and sustaining app use over time. These results match the observation that emotions matter for patients' engagement in earlier studies~\cite{gizaw_what_2022,yang_digital_2024,thomas_methods_2008,niyomyart_current_2024}. In future design, caregivers and developers should prioritise life quality as the key motivation for elderly users when adopting digital health aged-care solutions. In doing so, developers can help shift mHealth from a purely functional tool to one that promotes autonomy, purpose, and emotional well-being—factors that ultimately sustain engagement in ageing populations.
% Uncomment and use as the case may be
%\begin{theorem} 
%\end{theorem}

% Uncomment and use as the case may be
%\begin{lemma} 
%\end{lemma}

\section{Threats to Validity}\label{Threats to Validity}

This study has several limitations. First, a primary limitation of this study is its cross-sectional design, which captures stakeholder perspectives at a single point in time. Furthermore, the survey relied on self-reported data, which may have been influenced by subjective interpretation despite efforts to provide clear definitions and examples. Although surveys enable efficient data collection from large samples, they inherently lack the depth and contextual insight offered by qualitative methods such as interviews. Despite conducting pilot testing to ensure the clarity of survey items, potential misinterpretation of questions cannot be ruled out. Second, although the sample included developers, caregivers, and older adults from diverse backgrounds, it may not fully represent all aged-care digital health stakeholders globally. The recruitment strategy — relying on professional networks and social media — introduces a potential selection bias toward individuals who are more digitally literate. Participants in each group were independent, with no evidence of direct relationships (e.g., patient–caregiver or clinician–developer pairs), which may limit insights into real-world interdependent dynamics. Third, the survey did not include a validated health status measure, limiting our ability to examine how physical or cognitive health conditions moderate technology perceptions. Finally, while group comparisons were conducted, individual factors such as age, health, and caregiving experience were not analysed in depth. To address these issues, the study incorporated pilot testing, clear item design, and quantitative–qualitative triangulation, but future longitudinal or mixed-method research is needed to explore how human and relational factors shape aged-care software use over time.

\section{Conclusions}\label{conclusions}

This study presents the first empirical three-group comparison of requirements perceptions across older adults, caregivers, and developers in aged-care digital health, identifying current satisfaction, limitations, and future desired innovations for functional and non-functional requirements in digital health for older adults. We validated requirements reported in a prior systematic review and extended them through open-ended survey responses. Using both quantitative and qualitative data from older adults, caregivers, and developers, we examined where stakeholder priorities converge and where they diverge across current systems and future desired designs.
\\
The results show that functional requirements are generally rated more positively than non-functional requirements, and that older adults and caregivers are often more positive than developers. However, older adults also expressed more critical views in several areas, challenging the assumption that core usability and accessibility needs are already sufficiently addressed. Our stakeholder-gap analysis further shows that these gaps are patterned rather than random: developers are often more distant from end-user groups, especially for current functional requirements, while caregivers, particularly informal caregivers, tend to express stronger expectations for future requirements.
\\
These findings suggest that digital health design for aged care should not rely on any single stakeholder perspective. Instead, future development should concentrate on the most critical functional requirements, prioritise the non-functional qualities that shape usability and adherence, especially simplicity, responsiveness, and seamless integration, and address human factors more directly, including health-related conditions, emotional wellbeing, socio-economic context, and the support available from formal and informal caregivers.
\\
%Overall, this study provides a stakeholder-gap analysis that can support co-design, requirements prioritisation, near-term product decisions, and privacy-by-design in aged-care digital health.

\section{Acknowledgements}\label{acknowledgements}
Xiao and Grundy are supported by ARC Laureate Fellowship FL190100035.
% To print the credit authorship contribution details
\printcredits
%\clearpage
%% Loading bibliography style file
% \bibliographystyle{model1-num-names}
\bibliographystyle{elsarticle-num}
% \bibliographystyle{cas-model2-names}

% Loading bibliography database
\bibliography{cas-refs}
%% The Appendices part is started with the command \appendix;
%% appendix sections are then done as normal sections
%\clearpage
\appendix
\section{Qualitative Coding of Open-Ended Survey Responses}
\label{app:coding}

The following figure illustrates the qualitative coding process applied to 
open-ended responses.

\includepdf[pages=-, scale=0.9]{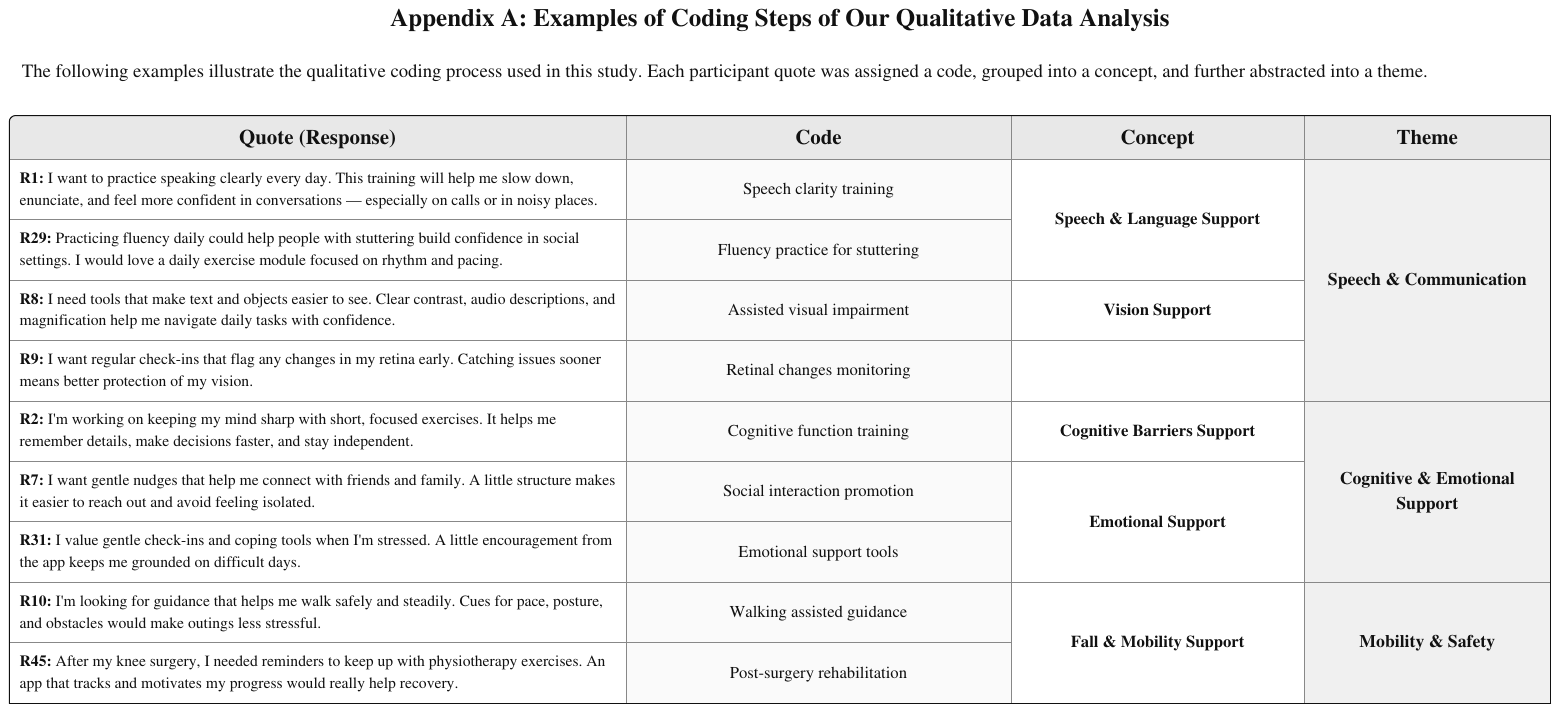}

\section{Quantitative Analysis of Likert-scale Survey Responses}

\begin{table*}[t]
\centering
\scriptsize
\setlength{\tabcolsep}{3pt}
\renewcommand{\arraystretch}{1.1}
\begin{threeparttable}
\caption{One-Way ANOVA Results for Perceived Importance of Functional Requirements Across Stakeholder Groups: Current vs.\ Future}
\label{tab:anova_functional_full}
\begin{tabular}{p{1.7cm}p{0.7cm}p{1.7cm}p{1.7cm}p{1.7cm}p{1.7cm}cccc}
\toprule
\textbf{Requirements}
  & \textbf{Type}\tnote{d}
  & \textbf{OA}\tnote{b}
  & \textbf{Dev}\tnote{b}
  & \textbf{IC}\tnote{b}
  & \textbf{FC}\tnote{b}
  & \textbf{F}
  & \textbf{$p$}
  & \textbf{$\eta^2$}
  & \textbf{Perception Order}\tnote{c} \\
\midrule
Chronic Dis.\ Mgmt
  & Current & 3.7 (3.36--3.97) & 3.0 (2.76--3.16) & 5.0 (5.00--5.00) & 4.0 (2.70--5.30)
  & 13.572.115  & $<$.001 & 0.199
  & Dev $\ll$ OA $<$ FC $\approx$ IC \\
  & Future  & 4.0 (3.80--4.18) & 3.7 (3.50--3.94) & 4.4 (3.70--5.16) & 2.0 (1.64--2.66)
  & 7.115  & $<$.001 & 0.085
  & Dev $\approx$ OA $<$ FC $\approx$ IC \\[4pt]
Fall/Mobility Supp.
  & Current & 3.3 (2.95--3.62) & 3.0 (2.85--3.13) & 4.5 (4.05--4.95) & 4.8 (3.95--5.55)
  & 9.666  & $<$.001 & 0.150
  & Dev $\ll$ OA $<$ FC $\approx$ IC \\
  & Future  & ---             & 3.8 (3.56--4.02) & 4.4 (3.70--5.16) & 4.8 (4.66--4.97)
  & 9.23  & $<$.001 & 0.180
  & Dev $<$ OA $<$ IC $<$ IC \\[4pt]
Life Assistance
  & Current & ---              & 3.0 (2.81--3.16) & 4.5 (4.05--4.95) & 4.5 (3.58--5.42)
  & 7.642  & $<$.001 & 0.123
  & Dev $\approx$ OA $<$ FC $=$ IC \\
  & Future  & 3.9 (3.72--4.08) & 3.8 (3.52--3.98) & 4.3 (3.26--5.31) & 4.7 (4.53--4.95)
  & 7.118  & $<$.001 & 0.085
  & Dev $\approx$ OA $<$ FC $\approx$ IC \\[6pt]
% %--- Unsatisfactory ---
% \multicolumn{10}{l}{\textit{Unsatisfactory functional requirements (RQ2.1 / RQ4.2)}} \\[2pt]
Med \& Safety
   & Current & 4.4 (4.11--4.72) & 3.0 (2.85--3.15) & 4.0 (3.11--4.89) & 4.3 (2.73--5.77)
  & 4.304  & .006    & 0.070
  & Dev $<$ OA $\approx$ FC $\approx$ IC \\
  & Future  & 4.0 (3.78--4.14) & 3.9 (3.73--4.15) & 4.6 (4.08--5.07) & 3.0 (2.50--3.52)
  & 6.734  & $<$.001 & 0.080
  & Dev $\approx$ OA $<$ FC $\approx$ IC \\[4pt]
Mental Health\tnote{a}
  & Current & ---              & 3.0 (2.50--3.52) & 4.0 (3.11--4.89) & 4.3 (3.45--5.05)
  & 229.6  & $<$.001 & 0.818
  & Dev $\ll$ FC $\approx$ IC \\
  & Future  & \multicolumn{7}{l}{\textit{(---)}} \\[4pt]
Brain Training
  & Current & 3.5 (3.24--3.81) & 2.8 (2.57--2.98) &  4.0 (3.11--4.89) & 4.3 (2.73--5.77)
  & 9.640  & $<$.001 & 0.150
  & Dev $<$ \{OA $\approx$ FC $\approx$ IC\} \\
  & Future  & 3.9 (3.70--4.06) & 3.9 (3.61--4.09) & 4.7 (4.53--4.95) & 4.6 (4.08--5.07)
  & 6.873  & $<$.001 & 0.082
  & Dev $\approx$ OA $<$ FC $\approx$ IC \\[4pt]
Fitness Support
  & Current & 3.6 (3.28--3.86) & 3.0 (2.86--3.10) & 4.0 (2.70--5.30) & 5.0 (5.00--5.00)
  & 18.600 & $<$.001 & 0.254
  & Dev $<$ OA $<$ FC $\approx$ IC \\
  & Future  & \multicolumn{7}{l}{\textit{(---)}} \\[4pt]
Rehab Support
  & Current & ---              & 2.7 (2.53--2.91) & 4.0 (3.11--4.89) & 4.3 (3.45--5.05)
  & 15.420 & $<$.001 & 0.220
  & Dev $<$ \{OA $\approx$ FC $\approx$ IC\} \\
  & Future  & 3.6 (3.33--3.78) & 3.8 (3.61--4.05) & 4.7 (4.26--5.17) & 3.0 (2.50--3.52)
  & 8.229  & $<$.001 & 0.096
  & Dev $\approx$ OA $<$ FC $\approx$ IC \\
\bottomrule
\end{tabular}
\begin{tablenotes}[flushleft]
\footnotesize
\item M = mean; 95\% CI = 95\% confidence interval (Lower--Upper Bound). future CIs are from SPSS Descriptives output. Effect size: small $\geq$.01, medium $\geq$.06, large $\geq$.14.
\item$F$ is the one-way ANOVA test statistic, computed as the ratio of between-group variance to within-group variance. Larger $F$ values indicate that differences among group means are larger relative to variability within groups.
\item$p$ is the significance level associated with the ANOVA test. A small $p$ value (typically $p<.05$) indicates that the observed group differences are unlikely to be due to random variation alone, suggesting a statistically significant difference among the stakeholder groups.
\item$\eta^2$ was calculated as $SS_{\text{between}} / SS_{\text{total}}$. SS: Sum of Squares of score in Anova test.
\item[a] --- means do not have enough answers to be sampled.
\item[b] OA = Older Adults; Dev = Developers; IC = Informal Caregivers; FC = Formal Caregivers. Mean value.
\item[c] \textit{Perception Order} reflects the ordering of group means based on post-hoc pairwise comparisons. Tukey HSD was applied when Levene's test indicated homogeneity of variance; Games-Howell was used when variances were unequal. '<<' denotes a highly significant difference (p < .001); '<' denotes a significant difference (p < .05). Rankings are intended to highlight perception gaps between stakeholder groups and inform design priorities for developers.
\item[d] Type of perception. Current = stakeholders' perceived satisfactory levels of existing systems; Future = stakeholders' desired levels for future systems.
\end{tablenotes}
\end{threeparttable}
\end{table*}
\begin{table*}[!ht]
\centering
\scriptsize
\setlength{\tabcolsep}{3pt}
\renewcommand{\arraystretch}{1.15}
\begin{threeparttable}
\caption{One-Way ANOVA Results for Non-Functional Requirements Across Stakeholder Groups: Current vs.\ Future}
\label{tab:anova_nonfunc_full}
\begin{tabular}{p{1.7cm}p{0.7cm}p{1.7cm}p{1.7cm}p{1.7cm}p{1.7cm}cccc}
\toprule
\textbf{Requirements}
  & \textbf{Type}\tnote{d}
  & \textbf{OA}\tnote{b}
  & \textbf{Dev}\tnote{b}
  & \textbf{IC}\tnote{b}
  & \textbf{FC}\tnote{b}
  & \textbf{F}
  & \textbf{$p$}
  & \textbf{$\eta^2$}
  & \textbf{Perception Order}\tnote{c} \\
\midrule
Entertainment/Social
  & Current & 3.7 (3.52--3.98) & 4.0 (3.80--4.22) & 4.7 (4.45--4.90) & 4.0 (2.16--5.84)
  & 0.29 & .750 & .008
  & OA $\approx$ Dev $\approx$ FC $\approx$ IC \\
  & Future  & 3.9 (3.78--4.10) & 4.0 (3.80--4.22) & 3.5 (3.05--3.95) & 4.6 (4.46--4.84)
  & 23.455& $<$.001&0.213&IC $<$ OA $\approx$Dev  $<$ FC \\
Personalisation
  & Current & 4.0 (3.70--4.21) & 4.0 (3.81--4.13) & 4.1 (3.64--4.59) & 4.0 (2.70--5.30)
  & 0.16 & .853 & .005
  & OA $\approx$ Dev $\approx$ FC $\approx$ IC \\
  & Future  & 4.1 (3.86--4.24) & 3.8 (3.58--4.02) & 4.5 (4.05--4.95) & 4.3 (3.93--4.65)
  & 4.48 & .012 & .040
  & OA $\approx$ FC $<$ IC \\[4pt]
Quickly Responsive\tnote{a}
  & Current & 3.9 (3.65--4.10) & ---              & 4.7 (4.50--4.97) & 4.3 (2.73--5.77)
  & 2.13 & .127 & .059
  & OA $\approx$ FC $\approx$ IC \\
  & Future  & 3.8 (3.65--4.01) & 3.8 (3.52--3.98) & 5.0 (5.00--5.00) & 4.6 (4.41--4.89)
  & 33.22 & $<$.001 & .278
  & OA $<$ FC $<$ IC \\[6pt]
% %--- Unsatisfactory ---
% \multicolumn{10}{l}{\textit{Unsatisfactory non-functional requirements (RQ2.2 / RQ4.2)}} \\[2pt]
Simple to Use
  & Current & 3.8 (3.57--4.05) & 3.9 (3.67--4.12) & 4.8 (4.59--4.94) & 3.5 (2.91--6.09)
  & 1.29 & .278 & .012
  & OA $\approx$ Dev $\approx$ FC $\approx$ IC \\
  & Future  & 4.0 (3.78--4.16) & 3.9 (3.67--4.12) & 4.0 (4.00--4.00) & 4.7 (4.50--4.91)
  & 39.68 & $<$.001 & .185
  & OA $<$ FC $\approx$ IC \\[4pt]
Visual Accessibility\tnote{a}
  & Current & 3.8 (3.51--4.01) & 3.9 (3.66--4.09) & 3.5 (1.91--5.09) & 4.3 (2.73--5.77)
  & 1.17 & .320 & .164
  & FC $\approx$ IC\tnote{d} \\
  & Future  & ---              & 3.9 (3.66--4.09) & 5.0 (5.00--5.00) & 4.7 (4.50--4.91)
  & 23.61 & $<$.001 & .214
  & OA $<$ FC $<$ IC \\[4pt]
Seamless Integrated
  & Current & 3.7 (3.50--3.88) & 4.6 (4.41--4.89) & 4.6 (4.35--4.89) & 4.3 (2.25--6.25)
  & 0.94 & .394 & .027
  & OA $\approx$ Dev $\approx$ FC $\approx$ IC \\
  & Future  & 4.0 (3.78--4.16) & 3.8 (3.59--4.03) & 5.0 (5.00--5.00) & 4.4 (4.08--4.75)
  & 10.09 & $<$.001 & .104
  & OA $<$ FC $<$ IC \\[4pt]
Health Info \& Tutorial\tnote{a}
  & Current & 3.8 (3.55--4.04) & ---              & 4.7 (4.54--4.93) & 4.0 (2.70--5.30)
  & 0.51 & .600 & .015
  & OA $\approx$ FC $\approx$ IC \\
  & Future  & 4.0 (3.79--4.13) & 3.8 (3.57--4.03) & 5.0 (5.00--5.00) & 4.6 (4.32--4.86)  
  & 23.455& $<$.001&.213  & FC $<$ OA $\ll$ IC\\
\bottomrule
\end{tabular}
\begin{tablenotes}[flushleft]
\footnotesize
\item M = mean; 95\% CI = 95\% confidence interval (Lower--Upper Bound). future CIs are from SPSS Descriptives output. Effect size: small $\geq$.01, medium $\geq$.06, large $\geq$.14.
\item$F$ is the one-way ANOVA test statistic, computed as the ratio of between-group variance to within-group variance. Larger $F$ values indicate that differences among group means are larger relative to variability within groups.
\item$p$ is the significance level associated with the ANOVA test. A small $p$ value (typically $p<.05$) indicates that the observed group differences are unlikely to be due to random variation alone, suggesting a statistically significant difference among the stakeholder groups.
\item$\eta^2$ was calculated as $SS_{\text{between}} / SS_{\text{total}}$. SS: Sum of Squares of score in Anova test.
\item[a] --- means do not have enough answers to be sampled.
\item[b] OA = Older Adults; Dev = Developers; IC = Informal Caregivers; FC = Formal Caregivers. Mean value.
\item[c] \textit{Perception Order} reflects the ordering of group means based on post-hoc pairwise comparisons. Tukey HSD was applied when Levene's test indicated homogeneity of variance; Games-Howell was used when variances were unequal. '<<' denotes a highly significant difference (p < .001); '<' denotes a significant difference (p < .05). Rankings are intended to highlight perception gaps between stakeholder groups and inform design priorities for developers.
\item[d] Type of perception. Current = stakeholders' perceived satisfactory levels of existing systems; Future = stakeholders' desired levels for future systems. 
\end{tablenotes}
\end{threeparttable}
\end{table*}
% % \include{appendices/appendixB}
% % Biography
% %\bio{}
% % Here goes the biography details.
% %\endbio

% %\bio{pic1}
% % Here goes the biography details.
% %\endbio

\end{document}